\documentclass[footinbib,a4paper,aps,prx,superscriptaddress,reprint,twocolumn,preprintnumbers,amsmath,amssymb,nobalancelastpage,10pt]{revtex4-1}
\usepackage{amsmath}
\usepackage{verbatim}
\usepackage{graphicx}
\usepackage{color}
\usepackage{tikz}
\usepackage{siunitx}
\usepackage{physics}
\usepackage{subfigure}
\usepackage{float}

\usepackage{mathptmx}
\usepackage{amssymb}
\usepackage{amsmath}
\usepackage{amsfonts}
\usepackage{array}

\usepackage{bm}
\usepackage{xcolor}
\usepackage{braket}

\definecolor{mygrey}{gray}{0.35}
\definecolor{myblue}{rgb}{0.2,0.2,0.8}
\definecolor{myzard}{cmyk}{0,0,0.05,0}
\definecolor{mywhite}{rgb}{1,1,1}
\definecolor{myred}{rgb}{1,0.,0.3}

\usepackage[colorlinks=true,citecolor=myblue,linkcolor=myred]{hyperref}

\def\half{\textstyle\frac{1}{2}}

\def\beq{\begin{equation}}
\def\eeq{\end{equation}}

\def\barray{\begin{eqnarray}}
\def\earray{\end{eqnarray}}

\begin{document}


\title{ $\mathbb{Z}_n$ solitons in intertwined topological phases}

\author{D. Gonz\'{a}lez-Cuadra}\email{daniel.gonzalez@icfo.eu}
\affiliation{ICFO - Institut de Ci\`encies Fot\`oniques, The Barcelona Institute of Science and Technology, Av. Carl Friedrich Gauss 3, 08860 Castelldefels (Barcelona), Spain}

\author{A. Dauphin}
\affiliation{ICFO - Institut de Ci\`encies Fot\`oniques, The Barcelona Institute of Science and Technology, Av. Carl Friedrich Gauss 3, 08860 Castelldefels (Barcelona), Spain}

\author{P. R. Grzybowski}
\affiliation{Faculty of Physics, Adam Mickiewicz University, Umultowska 85, 61-614 Pozna{\'n}, Poland}
\author{M. Lewenstein}
\affiliation{ICFO - Institut de Ci\`encies Fot\`oniques, The Barcelona Institute of Science and Technology, Av. Carl Friedrich Gauss 3, 08860 Castelldefels (Barcelona), Spain} 
\affiliation{ICREA, Lluis Companys 23, 08010 Barcelona, Spain}

\author{A. Bermudez}
\affiliation{Departamento de F\'{i}sica Te\'{o}rica, Universidad Complutense, 28040 Madrid, Spain}

\begin{abstract}

Topological phases of matter can support fractionalized quasi-particles localized at topological defects. The current understanding of these exotic excitations, based on the celebrated bulk-defect correspondence, typically  relies on crude approximations  where such defects are replaced  by a static  classical background  coupled to the  matter sector. In this work, we explore the  strongly-correlated nature of symmetry-protected topological defects by focusing on situations where such  defects arise spontaneously as dynamical solitons in intertwined topological phases, where  symmetry breaking coexists with topological symmetry protection.
In particular, we focus on the $\mathbb{Z}_2$ Bose-Hubbard model, a one-dimensional chain of interacting bosons coupled to $\mathbb{Z}_2$ fields, and show how solitons with $\mathbb{Z}_n$ topological charges appear for  particle/hole dopings  about certain  commensurate fillings, extending the results of \cite{short_paper} beyond half filling. We show that these defects host fractionalized bosonic quasi-particles, forming bound states that  travel through the system unless externally pinned, and repel each other  giving rise to a fractional soliton lattice for sufficiently high densities. Moreover, we uncover the topological origin of these fractional bound excitations through a pumping mechanism, where the quantization of the inter-soliton transport allows us to establish a generalized bulk-defect correspondence. This in-depth analysis of dynamical topological defects bound to fractionalized quasi-particles, together with the possibility of implementing our model in cold-atomic experiments, paves the way for further exploration of exotic topological phenomena in strongly-correlated systems.

\end{abstract}

\maketitle
\setcounter{tocdepth}{2}
\begingroup
\hypersetup{linkcolor=black}
\tableofcontents
\endgroup

\maketitle

\section{\bf Introduction}

Solitons are  waves that, despite arising in dispersive media,  propagate  with a well-defined velocity  and an unperturbed  shape due to  non-linear effects~\cite{PhysRevLett.15.240}. These solitary waves, first observed in  hydrodynamics~\cite{russell},  have fascinated scientists ever since, with premieres in  areas as diverse as biology~\cite{DAVYDOV1977379}, optics~\cite{solitons_fibers},  high-energy physics~\cite{skyrme}  and condensed matter~\cite{abrikosov}. It is in these two later disciplines where, despite the vast difference in energy scales, the origin of solitons has a common ground: the interplay of symmetry and topology~\cite{Finkelstein,toulouse}. 

The importance of symmetry in modern physics can be hardly exaggerated~\cite{Gross14256}. Invariance with respect to global/local symmetries has been the key  principle to find the fundamental laws of nature. Additionally, the spontaneous breaking of those symmetries is paramount to understand the emergence of  a wide variety of  phenomena   from the same  microscopic laws~\cite{Anderson393}. For instance, different phases of matter  can be characterized by the values of a local order parameter,  consistent with the spontaneous symmetry breaking (SSB) mechanism~\cite{landau}. In certain situations,  such as during  phase transitions~\cite{kibble,zurek}, the order parameter  may adopt  inhomogeneous configurations   interpolating between those different values. This gives rise to a  freely-propagating localized excitation that cannot be removed by smooth local deformations, nor understood using perturbation theory: a {\it topological soliton}.

 Solitons,  also  known  as  defects in this context,  are  relics of the original disordered phase  that distort  the  symmetry-broken  groundstate around a certain  core/center where the order parameter vanishes. Moreover, the winding of the order parameter around such defect cores  yields a topological invariant, underlying the  topological characterization of such defects. These winding numbers can only  change via non-local  deformations,   guaranteeing the robustness of the soliton    to physical,  primarily local, perturbations~\cite{RevModPhys.51.591}. Let us remark that  topological solitons are finite-energy non-perturbative solutions of classical field equations. Despite the fact that they can be  quantized formally~\cite{RevModPhys.49.681}, the interesting interplay of symmetry and topology is, in this case, a classical feature. 
 
Genuine quantum effects can become manifest when these solitons interact with quantum matter.
 As first predicted in relativistic quantum field theories~\cite{relativistic_solitons,jackiw_rossi} and, independently, in linear conjugated polymers~\cite{ssh} and $p$-wave superconductors~\cite{read_green}, topological solitons can bind quasi-particles with a fractional  numbers and exotic quantum statistics. As such, these {\it bound fractionalized quasi-particles} cannot be  adiabatically connected to the original particle content of the theory, as  typically occurs in  more standard situations such as Fermi liquids~\cite{landau_fl} or, more generally, perturbatively-renormalised  quantum field theories~\cite{WILSON197475}. Let us note that, in contrast to the topological soliton, these quasi-particles are not necessarily protected by any quantized topological  invariant. 
 
 With the  advent of topological insulators~\cite{PhysRevLett.95.146802,Bernevig1757,Konig766,Hsieh2008}, however, a quantum-mechanical protection mechanism  has been unveiled~\cite{topological_defects_1}, giving rise to the concept of  {\it symmetry-protected topological defects}~(SPT-d). Under certain symmetry constraints, the  characterization of the bulk gapped matter  comprised in between two of these solitons   requires  yet another  topological invariant. In this case, the relevant topological invariant is no longer the winding number of a classical field, such as the order parameter, but  is instead determined by the Berry connection of the matter sector via the quantum-mechanical wavefunction.  We note that these invariants are quantized and cannot change under external symmetry-preserving perturbations, unless these perturbations suffice to close the  bulk energy gap, inducing a  quantum phase transition where the protecting symmetry is spontaneously broken. Moreover, the so-called  {\it bulk-defect correspondence} connects this bulk topological invariant to the   fractional quasi-particles bound to the defect,   and justifies the protection of these  quasi-particles with respect to symmetry-preserving perturbations.

\begin{figure*}[t]
  \centering
  \includegraphics[width=1\linewidth]{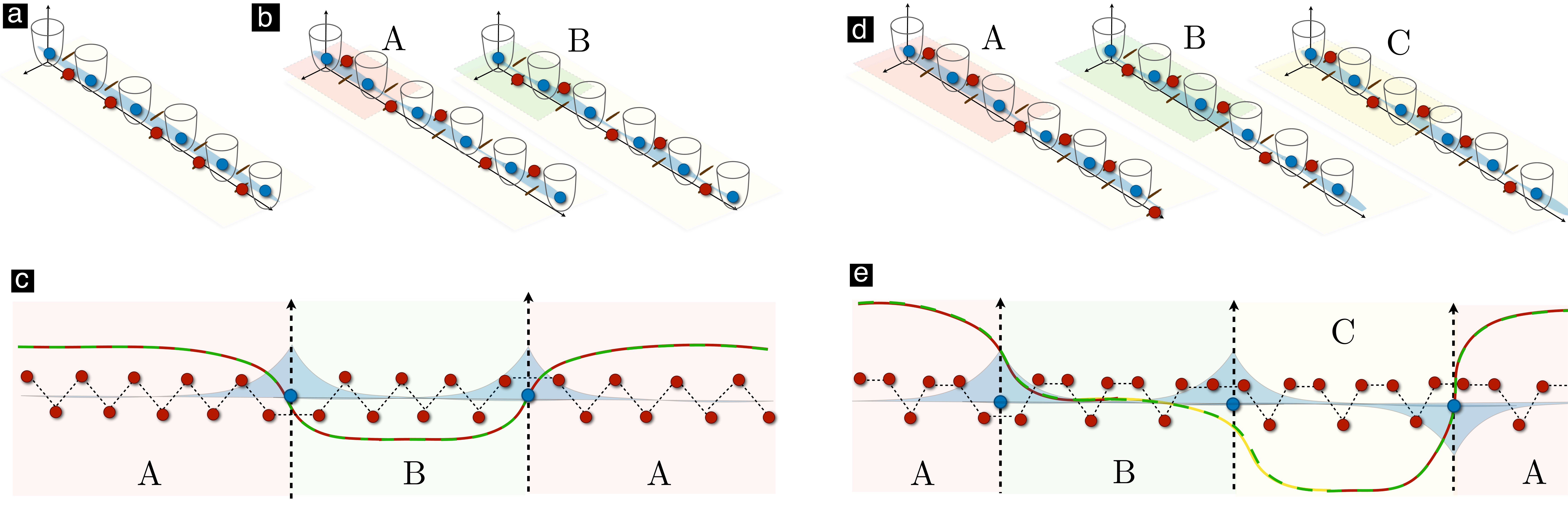}
\caption{\label{fig:domain_walls} {\bf Topological solitons in the $\mathbb{Z}_2$ Bose-Hubbard model:} Bosonic particles (blue spheres) can tunnel between neighboring sites of a chain, and their tunneling depends on the state of the $\mathbb{Z}_2$ fields (red spheres) at the bonds.  {\bf (a)} For weak Hubbard interactions, the groundstate of a half-filled system  ($\rho = 1 / 2$) is described by a bosonic quasi-superfluid delocalized over the chain, and a polarized background where all $\mathbb{Z}_2$ point in the same direction.  {\bf (b)} At stronger interactions, the $\mathbb{Z}_2$ fields order anti-ferromagnetically according to two degenerate patterns (A and B), spontaneously breaking translational invariance. Simultaneously, the bosons display a long-range order with an alternating bond density. Note that, in addition to the bond order wave,  one of the SSB sectors (B) hosts a symmetry-protected topological phase.  {\bf (c)} As a consequence of the SSB, the $\mathbb{Z}_2$ fields may adopt an inhomogeneous configuration with  topological solitons, such as the ABA background hereby displayed. As a consequence, bosonic quasi-particles are bound to the solitons. {\bf (d)-(e)} Analogue situation for the  trimer bond-ordered waves phases (valid for $\rho = 1 / 3$ and $\rho = 2 / 3$). In this case, the $\mathbb{Z}_2$ fields order ferri-magnetically according to three possible patterns (A, B and C), and a richer variety of topological solitons with bound quasi-particles  can form.}
\end{figure*}

The current theory of SPT-d  typically assumes a   background solitonic profile  that is static and externally fixed, focussing on the properties of the matter sector and the robustness of the bound quasi-particles~\cite{topological_defects_2}. This simplifies the description, as one  deals with non-interacting matter in an inhomogeneous classical background, allowing one to  identify the  mechanism responsible for their protection, and even to classify all possible SPT-d according to the underlying global/crystalline symmetries~\cite{classification_spt}.  Let us emphasize that, in this limit,  there is no intrinsic soliton dynamics, nor any SSB order parameter that would justify the topological protection of the soliton. Therefore, by assuming/engineering such externally-adjusted solitonic profiles, one is  missing half of the topological robustness  of the SPT-d. Moreover, from a fundamental point of view, this approach misses a key property: SPT-d can only arise in {\it intertwined topological phases}, which simultaneously display SSB and topological symmetry protection. Although one typically finds claims in the literature about  the absence of any local order parameter characterizing the SPT-d, the reality is that the matter sector can also display long-range order as a consequence of the SSB. It is the possibility of simultaneously encompassing both SSB long-range order and topological symmetry protection, which makes these intertwined SPT-d so exotic and interesting.This ambivalent role of symmetry is a consequence of interactions and, by exploring the full  non-perturbative nature of SPT-d, as well as the back action of the   matter  on the semi-classical topological soliton,  we expect that exotic   many-body effects will be unveiled, and new avenues of research will be open.

To address this  challenge, one can  start by exploring simplified models that are   still  complex enough to capture the intricate nature of the problem. In particular, reducing dimensionality allows for  efficient numerical techniques that can be used to test the validity of analytical predictions, and confirm the existence of SPT-d and the interplay of SSB and topological symmetry protection. In this work, we explore the existence of SPT-d in the so-called $\mathbb{Z}_2$ Bose-Hubbard model ($\mathbb{Z}_2$BHM)~\cite{z2_bhm_1,z2_bhm_2,z2_bhm_3} (see Fig.~\ref{fig:domain_walls}). In addition to the standard Bose-Hubbard dynamics of a one-dimensional chain of interacting bosons~\cite{bose-hubbard}, the $\mathbb{Z}_2$BHM includes $\mathbb{Z}_2$ quantum fields (i.e. Ising spins) that sit on the lattice bonds   and  dress the tunnelling of bosons according to the following Hamiltonian
\begin{equation}
\label{eq:z2_bhm}
\begin{split}
H =& -\sum_i \!\left[ b_i^\dagger(t+\alpha\sigma^z_{i,i+1})b_{i+1}^{\phantom{\dagger}}+\text{H.c.}\right] + \frac{U}{2}\sum_i n_i(n_i-1)\\
&+\frac{\Delta}{2}\sum_i\sigma^z_{i,i+1}+\beta\sum_i\sigma^x_{i,i+1}.
\end{split}
\end{equation}
Here,  $b^{\phantom{\dagger}}_i$ ($b^\dagger_i$) is the bosonic annihilation (creation) operator at site $i\in\{1,\cdots, N\}$, and $n_i=b^\dagger_i b_i$ counts the number of bosons, such that the  interaction strength $U>0$ quantifies the repulsion between pairs of bosons residing at the same lattice site. Additionally, $\sigma^x_{i,i+1},\sigma^z_{i,i+1}$ are  Pauli operators describing the  $\mathbb{Z}_2$ field at the bond $(i,i+1)$, the fluctuations  of which are controlled by  the transverse field  $\beta$. This model is reminiscent of a lattice gauge theory~\cite{RevModPhys.51.659}, where the  tunneling of bosons gets dressed with a relative strength $\alpha$ by  the $\mathbb{Z}_2$ fields. However, we emphasize that the appearance of the bare boson tunnelling  $t$, and the bare spin energy   $\Delta$, explicitly breaks the local $\mathbb{Z}_2$ gauge symmetry. We stress that in the rest of the work,  we fix $\alpha = 0.5\, t$, and  measure the energies in units  of $t$.
 
 As discussed in~\cite{z2_bhm_1,z2_bhm_2,z2_bhm_3}, the $\mathbb{Z}_2$BHM~\eqref{eq:z2_bhm} gives rise to  a bosonic version of the so-called Peierls dielectrics with  interesting topological features. Due to the Peierls' mechanism~\cite{peierls}, one-dimensional (1D) metals are known to be unstable towards a  dielectric insulator, where  a periodic lattice distortion develops in combination with  a charge/bond
 density wave. As shown in~\cite{z2_bhm_1,z2_bhm_2}, despite lacking a Fermi surface, the   $\mathbb{Z}_2$BHM   hosts similar Peierls' phenomena with interesting peculiarities: {\it (i)} the bosonic 1D system is not always unstable, but allows for a rich phase diagram. In particular, one finds a bosonic quasi-superfluid  in a polarized  $\mathbb{Z}_2$-field background [see Fig.~\ref{fig:domain_walls}{\bf (a)}], which is transformed at  strong interactions into a bond-ordered wave in two possible  N\'eel-ordered backgrounds (A- or B-type, as displayed in Fig.~\ref{fig:domain_walls}{\bf (b)}). {\it (ii)} At half-filling, the bond-ordered wave in the  B-type SSB sector is a neat example of an intertwined topological phase, which displays both long-range order and a non-zero topological invariant signalling the existence of edge states. This  topological bond-ordered wave (TBOW$_{1/2}$) is a bosonic counterpart of  certain  Peierls insulators, which are protected by inversion symmetry, and described by the so-called Su-Schrieffer-Hegger (SSH) model~\cite{ssh}.  {\it (iii)} In contrast to the fermionic  SSH model, which supports a TBOW$_{1/2}$  that can be adiabatically connected to a free-fermion SPT phase, the bosonic  TBOW$_{1/2}$ of the   $\mathbb{Z}_2$BHM requires sufficiently-strong inter-particle interactions, and cannot be adiabatically connected to a free-boson SPT phase even in the limit of a static lattice. It is thus a clear example of  an interaction-induced SPT phase. {\it (iv)} When considering other fractional fillings [see Fig.~\ref{fig:domain_walls}{\bf (d)}], further distinctions with respect to the fermionic case arise. For one-third (two-third) filling~\cite{z2_bhm_3}, the groundstate develops a TBOW$_{1/3}$ (TBOW$_{2/3}$) with no  counterpart in the SSH model~\cite{fractionalization_trimer,PhysRevB.27.370}, which  only hosts non-topological bond-ordered waves with the analogue of the A-, B- or C-type  orderings of  Fig.~\ref{fig:domain_walls}{\bf (d)}. As shown in~\cite{z2_bhm_3}, the $\mathbb{Z}_2$BHM  has a different region in parameter space where the three-fold degenerate groundstate displays  $\bar{\rm A}$-, $\bar{\rm B}$- or $\bar{\rm C}$-type ferrimagnetic ordering of the $\mathbb{Z}_2$ fields, obtained by a global  $\mathbb{Z}_2$  inversion of the corresponding A, B or C ferrimagnets of  Fig.~\ref{fig:domain_walls}{\bf (d)}. For such backgrounds,  the bosonic sector is described by a topological bond-ordered wave. In contrast to the half-filled case, where the SSB of translation invariance leads to a 2-site unit cell, and directly imposes inversion symmetry as the protection mechanism of the intertwined topological phase; the 3-site unit cell of one-third or two-third fillings does not necessarily impose the inversion symmetry. Instead, this symmetry protecting  the intertwined TBOW$_{1/3}$ (TBOW$_{2/3}$) does  emerge  at low energies and becomes responsible for various exotic effects, such as a fractional topological pumping.

The phenomena discussed above are driven by the interplay of SSB and topological symmetry protection, both of which occur simultaneously in the groundstate of the $\mathbb{Z}_2$BHM~\eqref{eq:z2_bhm} at commensurate fillings. However, as mentioned in the introduction, SSB also allows for the existence of topological defects/solitons and SPT-d. 
Given the pioneering results on   solitons in the fermionic SSH model~\cite{ssh_review}, and the number of similarities discussed in the previous paragraphs, it is a fair  question to assess if  topological solitons and fractionalization of bosons could be observed in systems described by the  $\mathbb{Z}_2$BHM. This question is even more compelling given the possibility of implementing this model using ultracold bosonic atoms in periodically-modulated optical lattices, following the lines of~\cite{Bermudez_2015,dd_fields_eth,z2LGT_1,z2LGT_2}. As substantiated in the following sections, this implementation would allow for a real breakthrough in the field: the first direct experimental observation of  fractionalization of matter by a non-static  soliton.

 The soliton model of SSH has been argued to play a key role in  the physics of a linear conjugate polymer, namely polyacetylene~\cite{ssh_review}. Despite   indirect  evidence,  there  has always been a certain degree of controversy about the role of solitons in the  properties of polyacetylene~\cite{laughlin_fractional}: long-range dimerized order, let alone a solitonic configuration,  has never been   observed directly. Moreover, the spin-full character of electrons in polyacetylene, which is half-filled, masks the  fractionalization. Despite  leading to reversed spin-charge relations, which  yield an  indirect evidence~\cite{ssh_review}, this spin doubling of polyacetylene  forbids a direct experimental confirmation of the bound fractionalized nature of  quasi-particles.  Accordingly, the existence of topological solitons and bosonic fractionalization shown below, together with the possible implementation of the $\mathbb{Z}_2$BHM  in experiments of ultra-cold atoms, opens a new  promising route in the  study of SPT-d.
 
In this work we summarize and extend the results of \cite{short_paper}, where the $\mathbb{Z}_2$ Bose-Hubbard model was studied for incommensurate densities close to half filling, and characterized the different symmetry-protected topological defects that appear for various other densities. The paper is organized as follows. In Sec.~\ref{sec:topological_solitons}, we show how topological solitons appear spontaneously in the groundstate of the  $\mathbb{Z}_2$BHM when it is doped above/below certain commensurate fillings, and we characterize them in terms of topological charges associated to the underlying SSB sectors. Moreover, we demonstrate how these defects bind  fractionalized bosonic quasi-particles. These composite objects can propagate through the chain, repelling each other at short distances. For a finite density of defects, this interaction gives rise to a fractional soliton lattice. In Sec.~\ref{sec:bulk-defect}, we explore how the topological properties of the matter sector can bring extra protection to these bound quasi-particles. In particular, we characterize the different SSB sectors as intertwined topological phases using symmetry-protected topological invariants. This allows one to track the origin of fractionalization in the system through a bulk-defect correspondence. Remarkably, the later can be generalized to situations where the regions separated by the defects are in the same topological sector. This requires extending the system to two dimensions using a pumping mechanism, where the 2D topological invariant, the so-called Chern number, can be  recovered by measuring the quantized inter-soliton transport. The conclusions and outlook are presented in Sec.~\ref{sec:conclusions}.

\section{\bf Topological solitons and boson fractionalization}
\label{sec:topological_solitons}

\subsection{$\mathbb{Z}_n$ solitons: doping and  pinning}

As advanced previously, topological solitons are stable finite-energy  excitations that may arise  when different values of the  order parameter  are allowed by SSB~\cite{Rajaraman:1982is}.  These excitations can be dynamically generated by crossing a symmetry-breaking critical point in a finite time. In this way, the ordered phase gets distorted by such   solitons, the density of which  scales with the crossing rate according to the so-called Kibble-Zurek scaling~\cite{kibble,zurek}. Let us remark that, once the phase transition has been crossed,  an extensive number of excitations are present in the system. These solitons  evolve in time, scattering off each other, or escaping through the edges of the system, which gives rise to   a complex out-of-equilibrium problem. In this subsection, we argue that the $\mathbb{Z}_2$BHM~\eqref{eq:z2_bhm}  can  host  solitons in the groundstate, and that they can be pinned externally, leading to a simpler equilibrium situation.

The crucial condition to find  solitons  directly in the groundstate is to allow for their coupling with matter, which turns the situation into a  very interesting  quantum many-body problem. This is predicted to occur in the SSH model of polyacetylene by doping above/below half-filling~\cite{SS_soliton_dynamics, Brazovskii,CB_solitons_GN} and, recently, also in a system of fermionic atoms inside an optical waveguide \cite{Fraser_2019}. As shown below, solitons also appear  in the $\mathbb{Z}_2$BHM~\eqref{eq:z2_bhm}, with characteristic differences. For instance,  a single boson above half-filling can fractionalize,  giving rise to two  quasi-particles bound to a soliton-antisoliton pair in the groundstate (see the qualitative scheme in Fig.~\ref{fig:domain_walls}{\bf (c)}). For one-third and two-third fillings, a single boson can give rise to a  richer profile of topological solitons and bound fractional quasi-particles [see the qualitative scheme  in Fig.~\ref{fig:domain_walls}{\bf (e)}]. One of the key differences with respect to the soliton model of polyacetylene is that, during the doping process in the polymer,  all sorts of additional disorder and randomness are inebitably introduced~\cite{KIVELSON200199}. This disorder is likely the underlying source of  difficulties in providing an unambiguous proof of the dimerised long-range order and the associated  solitonic profiles in polyacetylene~\cite{laughlin_fractional}. In this context, an advantage of    the  $\mathbb{Z}_2$BHM~\eqref{eq:z2_bhm}, and its potential cold-atom relisation,  is that the atomic filling does not introduce impurities. In this way, one gets access to the groundstate of the doped system in a pristine environment, where the existence of solitons/fractionalization is not masked by  uncontrolled disorder. 

Let us now present a more systematic study of  topological solitons in the  $\mathbb{Z}_2$BHM, and provide quantitative evidence of the correctness of Figs.~\ref{fig:domain_walls}{\bf (c)} and {\bf (e)}. In the introduction, we presented solitons as localized finite-energy solutions moving at constant speed and, yet, Figs.~\ref{fig:domain_walls}{\bf (c)} and {\bf (e)}  represent static solitonic configurations. In a continuum  model of polyacetylene~\cite{PhysRevB.21.2388}, the solitons are indeed free to move. However, in a realistic situation,  charged impurities appear upon doping, and  play an additional important role: they pin the charged quasi-particles bound to the solitons at random positions~\cite{ssh_review}. In this work, we introduce a simple and deterministic pinning mechanism which, as shown below, allows us to  precisely control the position of the  solitons  and fractional bosons in the  $\mathbb{Z}_2$BHM. 
 
 Such a deterministic pinning is achieved by introducing a local perturbation of the $\mathbb{Z}_2$ fields, $H\to H+H_{\rm p}$, where 
 \beq
 \label{eq:pinning_term}
 H_{\rm p}=\sum_i\beta^{\phantom{x}}_i\sigma_{i,i+1}^x,\hspace{2ex} \beta^{\phantom{x}}_{i}=\sum_{j_{\rm p}\in{\rm P}}\beta\epsilon_{\phantom{i}} (\delta_{i-1,j_{\rm p}}+\delta_{i,j_{\rm p}}).
 \eeq
 Here,  $\epsilon$ stands for the relative strength of the pinning potential, and we sum over all  pinning centers labelled by $j_{\rm p}\in{\rm P}$. Essentially, this perturbation modifies the transverse field at  two  consecutive bonds that surround each of the pinning centers, $\beta\to\beta_0=\beta(1+\epsilon)$, selecting the corresponding solitonic profile that will depend on the particular filling.
 
 \subsubsection{Solitons with $\mathbb{Z}_2$-valued topological  charges}
\label{subsec:z2}

  Let us start by discussing the simplest situation, and address the appearance of topological solitons in the groundstate of the $\mathbb{Z}_2$BHM  doped above/below half-filling. As described above, there is a bosonic Peierls' transition where the Ising spins develop an antiferromagnetic  N\'eel-type order [see Fig.~\ref{fig:domain_walls}{\bf (b)}]. In this case, the order parameter can be defined as
 \beq
 \label{eq:Z2_order_parameter}
 \varphi=\frac{1}{N_{\rm u.c.}}\sum_{j=1}^{N_{\rm u.c.}}\varphi_j,\hspace{1ex}\varphi_j=\frac{1}{2}\sum_{i\in \text{u.c.}}\!\sin\left(\frac{\pi}{2}(2i-1)\right)\left\langle\sigma_{i,i+1}^z\right\rangle,
 \eeq
where $N_{\rm u.c.}$ is the number of unit cells (u.c.) labelled by $j$, and $\varphi_j$ is averaged over the elements of such unit cell $i\in{\rm u.c.}$ $\varphi$ is the total average staggered magnetization, and has two possible values $ \varphi=\pm \varphi_0\to\pm1$ depending on the SSB sector. Solitons will interpolate between these two vacua [See Fig.~\ref{fig:domain_walls}{\bf (c)}]. 

The existence of such solitons can be understood starting from the $\beta=0$ limit, where no quantum fluctuations exist, and they get reduced to  static domain walls.  Precisely at  half-filling, the  Peierls mechanisms results in antiferromagnetic N\'eel-like order the Ising spins. We recall that,  in the hard-core regime $U\rightarrow \infty$,  the Peierls transition is soleley controlled by $\Delta$. In this case, one can find analytical solutions~\cite{z2_bhm_1,z2_bhm_2} showing that N\'eel order coexists with a bosonic  bond-ordered wave  if $\Delta\in[\Delta_{\rm c}^{-},\Delta_{\rm c}^{+}]$, where $\Delta_{\rm c}^{\pm}=4t/\pi[\delta\pm (E(1-\delta^2)-1)]$ 
and we have introduced $\delta=\alpha/t$ and the complete elliptic integral of the second kind  $E(x)$. Conversely, for $\Delta\notin[\Delta_{\rm c}^{-},\Delta_{\rm c}^{+}]$, the  spins polarise along  the same direction  and the bosons form
  a quasi-superfluid state.

In this section, we explore a new situation that goes beyond the scope of these previous studies~\cite{z2_bhm_1,z2_bhm_2,z2_bhm_3}: we set $\Delta\in[\Delta_{\rm c}^{-},\Delta_{\rm c}^{+}]$, but explore lattices with an odd number of sites, such that perfect half-filling is  never possible. In this case, one can analytically show that the groundstate is not a single  N\'eel antiferromagnet in the Ising sector and  a dimerised bond-ordered wave for the bosons, but that it is composed of neighboring domains displaying the  possible SSB orders. In particular, we find that the groundstate can accommodate for a single domain wall connecting the two N\'eel states, and that this domain wall may come in two flavours:  ther are described by consecutive "up-up" or "down-down" spins  and, consecuently,  "strong-strong" (SS)  or "weak-weak"(WW) tunneling bonds for the bosons pattern respectively. The SS and WW domain walls couple differently to the itinerant bosons, and which one is stabilised in the groundstate  shall depend on the pinning and the particular number of sites.

 As $\beta$ is increased, we show below  that these domain walls widen into a soliton profile that coincides with the kink solutions of the (1+1) relatisvistic $\varphi^4$ theory~\cite{PhysRevD.10.4130}, namely
 \beq
 \label{eq:Z2_profile}
 \varphi_j=\tanh \left(\frac{j-j_{\rm p}}{\xi}\right),
 \eeq 
 where $j_{\rm P}$ is the soliton center and  $\xi$ is the width in lattice units. By analogy with the relativistic  scalar quantum field theory, one can define a topological charge~\cite{Rajaraman:1982is} as follows 
 \beq
 \label{eq:top_charge}
 Q= \half\left(\varphi_{j_{\rm p}+r}-\varphi_{j_{\rm p}-r}\right),
 \eeq
  where the order parameter is evaluated at points that are well separated from the soliton center, namely $r/\xi\to\infty$. In this case, the topological charge of the soliton is $Q=+1$, whereas anti-solitonic solutions carrying  $Q=-1$ can be obtained with the alternative interpolating profile  $ \varphi_j\to-\varphi_j$. These $\mathbb{Z}_2$  topological charges, which cannot be modified by local perturbations, guarantee the robustness  of the  soliton.

\begin{figure}[t]
\centering
\includegraphics[width=1.0\linewidth]{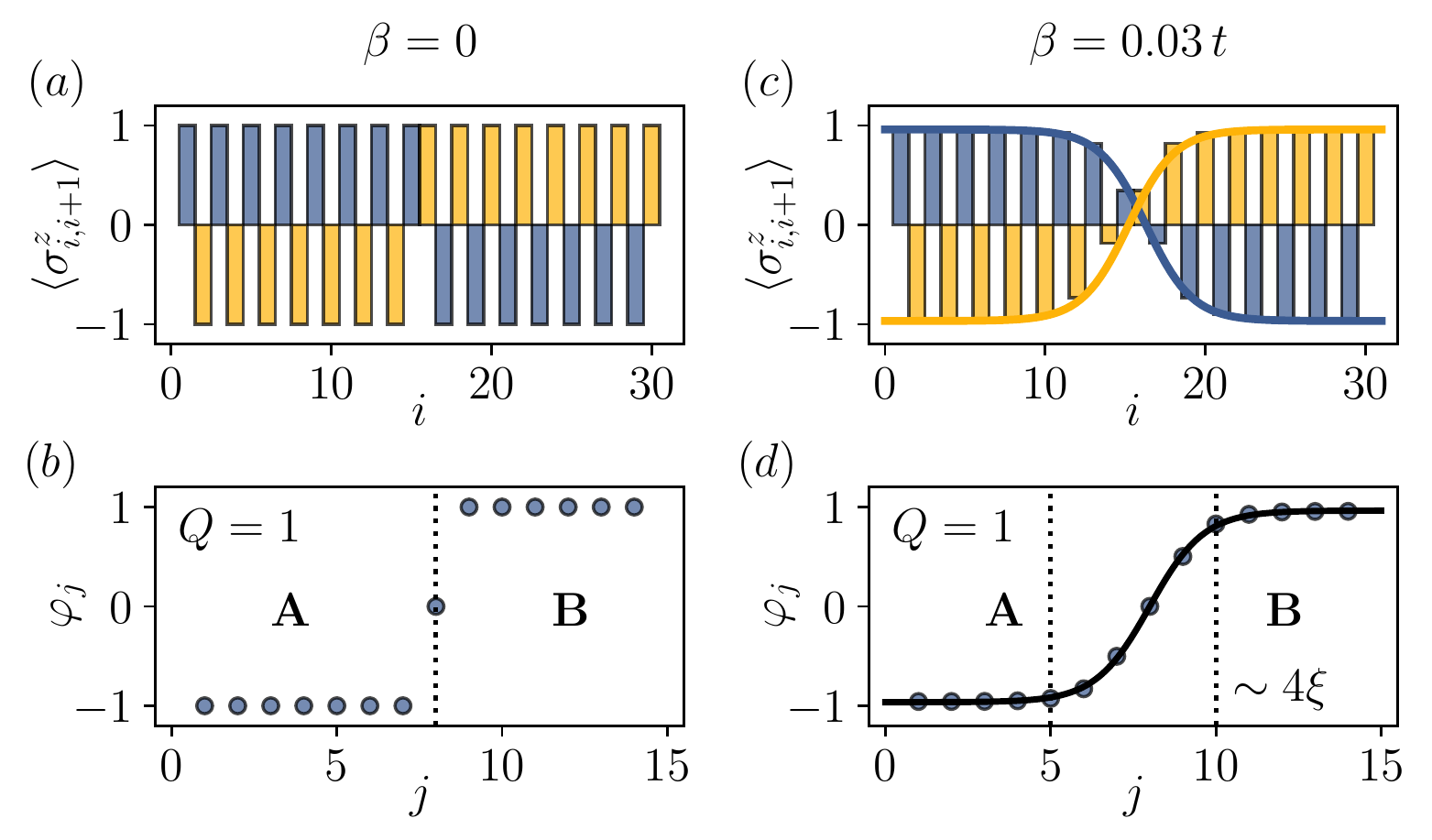}

\caption{\label{fig:loc_length} \textbf{$\mathbb{Z}_2$ topological defects: }{\bf (a)} Spin magnetization $\langle \sigma^z_{i, i+1} \rangle$ with a single topological defect modifying the dimerized pattern  of the BOW$_{1/2}$ phase, where different colours are used for the two sub-lattices. The defect corresponds to a domain wall ($\beta = 0$) connecting the two SSB sectors.  {\bf (b)} The defect interpolates between two values of the order parameter $\varphi_j$, which converges to $-1$ and $+1$ for $j=r-j_p$ and $j=r+j_p$, respectively, with $r \gg \xi$, and has a topological charge $Q = 1$. {\bf (c)-(d)} Analogous topological defect for  $\beta = 0.03t$, where we observe how quantum fluctuations broaden the defect, leading  to a soliton of finite width $\xi$. The order parameter, which was discontinuous for a domain wall, is smoothened for $\beta > 0$, and can be accurately fitted to Eq.~\eqref{eq:Z2_profile}. The parameters of the model are fixed to $U = 10\,t$ and $\Delta = 0.80\,t$, and we use a chain with $L=31$ sites and $N=16$ particles.}
\end{figure}

\begin{figure}[t]
\centering
\includegraphics[width=1.0\linewidth]{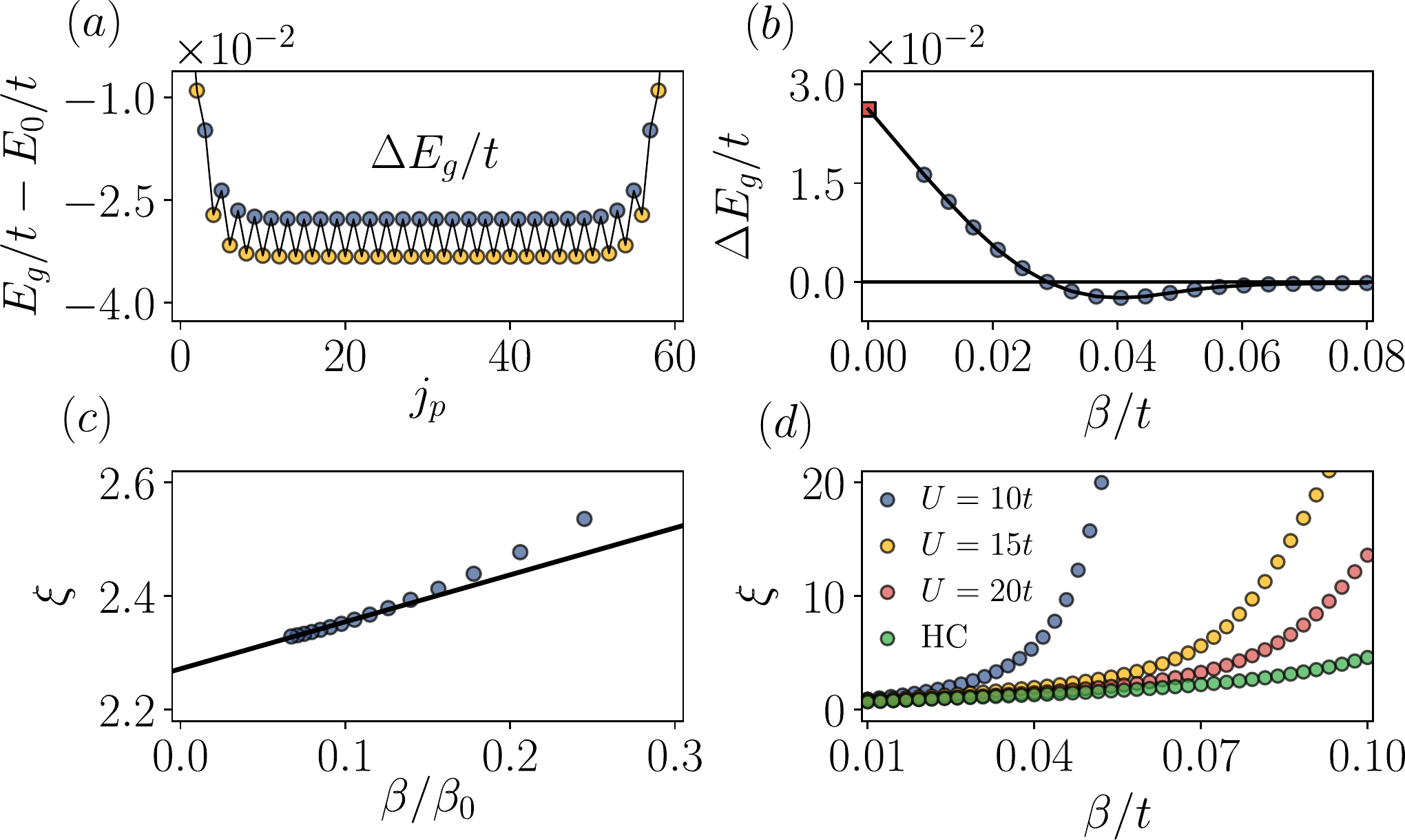}

\caption{\label{fig:loc_length_beta} \textbf{Effect of quantum fluctuations on topological defects: } {\bf (a)} Ground state energy $E_{\rm g}(j_{\rm p})$ with one topological defect as we vary the pinning location $j_{\rm p}$ on a chain of $L = 61$ sites, measured with respect to some reference $E_0$, for $\beta = 0.02\,t$. Odd and even sites are represented in different colors, showing the Peierls-Nabarro potential between them. Note that, far from the boundaries, the state is degenerate due to translational invariance. {\bf (b)} Peierls-Nabarro barrier $\Delta E_{\rm g}$ as a function of $\beta/t$, which goes to zero as the quantum fluctuations increase, for $L = 91$. For $\beta = 0$ we obtain a non-zero value for $\Delta E_g /t$ (red square). In this limit,  the defects are immobile. For $\beta\neq 0$ they can tunnel through the barriers. {\bf (c)} The pinning perturbation (Eq.~\eqref{eq:pinning_term}) localizes the defect at a certain position. Here we show how the soliton width  $\xi$ varies as we increase the pinning strength $\beta_0/\beta$, converging to a finite value for an infinite pinning. This finite-size scaling was performed by fitting the points to a line (black). In every panel, $\xi$ should be understood as the converged value. {\bf (d)} Soliton width  $\xi$ as a function of $\beta$ for different values of  the Hubbard interaction $U$, and we now work with  a chain of $L= 121$ sites. All the calculations where performed for one defect on top of the BOW$_{1/2}$ phase, with $U = 10\,t$ and $\Delta = 0.80\,t$.}
\end{figure}

To make contact with our previous analytical results, we note that the $\beta=0$ limit   of classical $\mathbb{Z}_2$ fields should yield an order parameter $ \varphi_j\sim \theta(j-j_{\rm p})$, where we have introduced the Heaviside step function. In this case, the topological soliton would have a vanishing width, and be localized within a single lattice site. To verify this prediction, we  study  the  $\mathbb{Z}_2$BHM on a chain of $L=31$ sites, filled with $N=16$ bosons. The ground-state has been obtained numerically using a density matrix renormalization group (DMRG) algorithm based on matrix product states (MPS) \cite{tenpy}. For the rest of the article, we use open boundary conditions and  bond dimension  $D = 100$. The maximum number of bosons per site is truncated to $n_0 = 2$, which is sufficient for strong interactions and low densities \cite{z2_bhm_1}. Figure~\ref{fig:loc_length}{\bf (a)} shows the  magnetization of the $\mathbb{Z}_2$ fields in the groundstate. As can be clearly observed, an SS domain wall is formed by  interpolating between the two possible anti-ferromagnetic  N\'eel  patterns. In this case, the defect is generated in the ground state since the odd number of sites does not allow for perfect half-filling,  but rather $\rho=N/L>1/2$. The existence of the domain wall becomes more transparent in Fig.~\ref{fig:loc_length}{\bf (b)}, where the corresponding order parameter~(\ref{eq:Z2_order_parameter}) displays the aforementioned step-like behavior.
 
Let us note that, in this classical limit, the solitons  can be centered around any lattice site $j_{\rm p}$ within the bulk of the system (i.e. the groundstate has an extensive degeneracy).  Moreover, in contrast to  continuum field theories where solitons are free to move, lattice solitons are static as they find finite energy barriers that inhibit their transport. These energy penalties, known as Peierls-Nabarro barriers, arise  due to the lack of translational invariance  on the lattice, and are typically arranged forming  a periodic  Peierls-Nabarro potential~\cite{Peierls_nabarro_1,Peierls_nabarro_2}. 
   
   As  the quantum fluctuations of the  $\mathbb{Z}_2$ fields are switched on, $\beta>0$, the topological solitons start tunnelling through these barriers, and delocalize along the  lattice. In order to study static soliton properties in this quantum-mechanical regime, it is important to switch on the pinning terms of Eq.~\eqref{eq:pinning_term}, which will effectively localize the soliton to the desired pinning center $j_{\rm p}$. However, as a result of quantum fluctuations, the soliton is no longer  strictly localized to a single site, but will  spread over a width  $\xi>0$.
   
    By externally modifying the position of the pinning center $j_{\rm p}$, we can confirm the existence of such Peierls-Nabarro barriers by numerically computing the ground-state energy as a function of the  position of the soliton $ E_{\rm g}(j_{\rm p})$ (see Fig.~\ref{fig:loc_length_beta}{\bf (a})). Due to the Peierls' dimerization, we find a periodic arrangement of energy  barriers $\Delta E_{\rm g}$ separating neighboring  unit cells. As shown in Fig.~\ref{fig:loc_length_beta}{\bf (b)}, the height of these barriers converges to a non-zero value when $\beta\to0$, leading to  the aforementioned Peierls-Nabarro barrier. This  numerical result justifies our previous statement about the static nature of the solitons in the classical regime: the finite barriers inhibit the movement of the topological solitons. The small magnitude of the Peierls-Nabarro  barrier also justifies our previous claim: as soon as quantum fluctuations are switched on, the solitons  tunnel   and delocalize, justifying the requirement of  pinning~\eqref{eq:pinning_term}. We note that the strength of the Peierls-Nabarro barriers may be larger for other choice of parameters, such that the tunnelling of solitons only gets activated at second-order in the transverse field $\beta^2$. In this particular regime, the very nature of the solitons is different, as they split into two distinct  branches (i.e. weak-weak and strong-strong bonds).  
   
   In  Fig.~\ref{fig:loc_length}{\bf (c)}, we represent such a pinned solitonic profile in presence of  quantum fluctuations. The pinning center is positioned at the middle $j_{\rm p}=(L-1)/2$ of a chain  of $L=31$ sites. As can be observed in Fig.~\ref{fig:loc_length}{\bf (d)}, the order parameter interpolates between the two SSB groundstates according to the smooth profile of Eq.~\eqref{eq:Z2_profile}, and an accurate fit can be used to extract the soliton width $\xi$. In Figure~\ref{fig:loc_length_beta}{\bf (c)}, we show how the soliton width gets modified as one increases the strength of the pinning potential,  eventually converging to a fixed value as the pinning strength is sufficiently strong $\beta/\beta_0=1/(1+\epsilon)\to 0$. Let us remark, however, that the modification of the soliton width for moderate pinnings   lies at the $10\%$ level, which is consistent with the contraction effect of other pinning mechanisms, such as the dopant impurities in the SSH model of polyacetylene~\cite{KIVELSON1986301}. In Figure~\ref{fig:loc_length_beta}{\bf (d)}, we represent the  widths $\xi$ of the  topological soliton as a function of the transverse field strength $\beta$, for different values of the Hubbard repulsion. We note that in all numerical simulations, we use  a pinning strength such that convergence of the soliton width has been reached. As expected,  the width increases (i.e. more delocalized defects) as  the quantum fluctuations are raised. On the other hand, we observe that the soliton width, at fixed $\beta$, decreases as the Hubbard repulsion $U$ is increased.  This is a clear demonstration of the back-action of the matter sector on the topological soliton that was briefly mentioned in the introduction: as the bosons become more repulsive, eventually reaching the hard-core constraint that forbids double occupancies, they can be accommodated  more comfortably within the $\mathbb{Z}_2$ soliton, which then becomes more localized.

   \subsubsection{Solitons with  $\mathbb{Z}_4$-valued  topological  charges}
   \label{subsec:z4}
   Let us note that, around half-filling, it is only possible to obtain two types of topological defects, solitons with charge $Q=+1$ and anti-solitons with  $Q=-1$. By doping, the groundstate configuration can only correspond to a succession of neighboring soliton and anti-solitons, such that the overall charge is either $Q_{\rm tot}=0$, or $Q_{\rm tot}=\pm1$ (i.e. even and odd topological sectors). For such configurations, one can imagine externally moving the defect centers by adjusting the pinning potentials, such that a soliton and anti-soliton scatter, annihilating each other. By repeating this process, one would reach a final situation where either one or none topological solitons remains. As  discussed below,  solitons and  possible scattering  are much richer around other fractional fillings.
 
 We shall now focus on  two-third filling, where the Peierls' mechanism can yield a three-fold degenerate ferrimagnetic ordering in the Ising sector [see Fig.~\ref{fig:domain_walls} {\bf (d)}]. The situation is analogous for filling one-third. The order parameter that can differentiate between these SSB groundstates is
 \beq
 \label{eq:Z4_order_parameter}
\tilde{\varphi}=\frac{1}{N_{\rm u.c.}}\sum_{j=1}^{N_{\rm u.c.}}\tilde{\varphi}_i,\hspace{2ex}\tilde{\varphi}_j=\frac{2}{\sqrt{3}}\sum_{i \in \text{u.c.}}\sin\left(\frac{2\pi}{3}(i-1)\right)\left\langle\sigma_{i,i+1}^z\right\rangle,
 \eeq
 which attains the following values for the the three  groundstates $ \tilde{\varphi}\in\{-2,0,2\}$. As there are more SSB vacua, there will be  more types of solitons that interpolate between them. For instance, an anti-soliton interpolating between the A and B groundstates (see Fig.~\ref{fig:domain_walls} {\bf (e)}) can be described as
  \beq
 \label{eq:AB_profile}
\tilde{ \varphi}^{\rm AB}_j=\left(1-\tanh \left(\frac{j-j_{\rm AB}}{\xi}\right)\right),
 \eeq 
 which has topological charge $Q_{\rm AB}=-1$ according to Eq.~\eqref{eq:top_charge}. The anti-soliton interpolating between the B and C   is
  \beq
 \label{eq:BC_profile}
\tilde{ \varphi}^{\rm BC}_j=\left(1+\tanh \left(\frac{j-j_{\rm BC}}{\xi}\right)\right),
 \eeq 
  which also has topological charge $Q_{\rm BC}=-1$. Finally, the soliton interpolating between C and A can be written as
   \beq
 \label{eq:CA_profile}
\tilde{ \varphi}^{\rm CA}_j=2\tanh \left(\frac{j-j_{\rm CA}}{\xi}\right),
 \eeq 
  which  has a larger topological charge $Q_{\rm CA}=+2$. We note that solitons with the inverse orderings BA, CB and AC, are also possible, and would lead to reversed topological charges, such that  solitons are restricted to a $\mathbb{Z}_4$-valued topological charge.
  
\begin{figure}[t]
\centering
\includegraphics[width=1.0\linewidth]{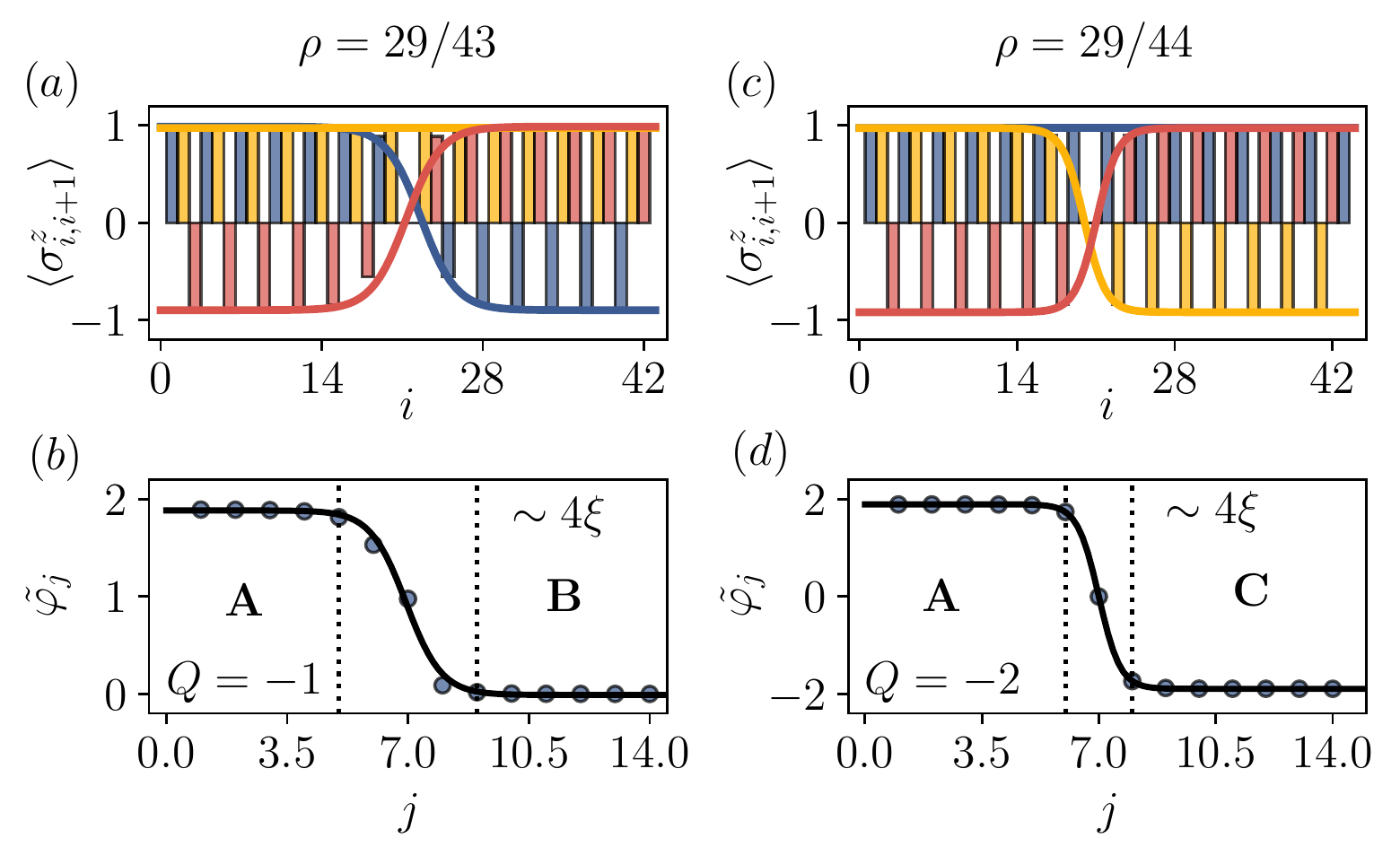}

\caption{\label{fig:loc_length_trimer} \textbf{$\mathbb{Z}_4$ topological defects: } {\bf (a)} Spin magnetization $\langle \sigma^z_{i, i+1}\rangle$ with a topological defect distorting  the trimerized pattern, and corresponding to an anti-soliton on the BOW$_{2/3}$ phase connecting two of the three SSB sectors, A and B, for a chain of $L = 43$ sites and $N=29$ particles. {\bf (b)} We represent the order parameter $\tilde{\varphi}_j$, where $j$ is the unit cell index. The black line is obtained by fitting $\tilde{\varphi}_j$ using eq.~\eqref{eq:AB_profile} and eq.~\eqref{eq:CA_profile}. Far from the defect, the order parameter distinguishes the different SSB sectors, and leads to an associated topological charge of $Q=-1$. The parameters of the model are fixed to $\beta = 0.01\, t$,  $\Delta = 0.85 \, t$ and $U = 10\,t$. {\bf (c)-(d)} Same as {\bf (a)-(b)}, but for an anti-soliton connecting the sectors A and C, for $L = 44$ and $N = 29$. This defect has an associated topological charge of $Q=-2$. }
\end{figure}
  
 In Figs.~\ref{fig:loc_length_trimer}{\bf (a)} and {\bf (c)} we provide  numerical confirmation of  this phenomenon, displaying two examples of defects connecting different SSB sectors of the BOW$_{2/3}$ phase. In both cases, the order parameter (\ref{eq:Z4_order_parameter}) adjusts very accurately to the expected shapes in Eqs.~\eqref{eq:AB_profile} and~\eqref{eq:BC_profile} [see Figs. \ref{fig:loc_length_trimer}{\bf (b)} and {\bf (d)}]. As it can be appreciated in the figure, different types of defects are generated depending on the bosonic density. In the next section we will analyze this mechanism in detail.   However, before turning into this discussion, let us emphasize that the wider variety of solitons hereby discussed leads to further possibilities regarding their scattering. In the ABCA sequence of Fig.~\ref{fig:domain_walls}{\bf (e)}, we see that this fractional filling  allows for a richer multi-solitonic profile with respect to the half-filling. In particular, one is no longer restricted to neighboring soliton-atisoliton pairs, but it becomes possible to find two neighboring defects with the same topological  charges $Q_{AB} = Q_{BC} = -1$. In this case, by externally adjusting the pinning potentials, these two defects can collide and lead to a larger conserved charge $Q_{AB} + Q_{BC} = -2$, which corresponds to a new type of anti- soliton AC with charge $Q_{\rm AC} = -2$ instead of the trivial vacuum, as occurred for half-filling.

\subsection{Boson  fractionalization}

\begin{figure*}[t]
  \centering
  \includegraphics[width=1\linewidth]{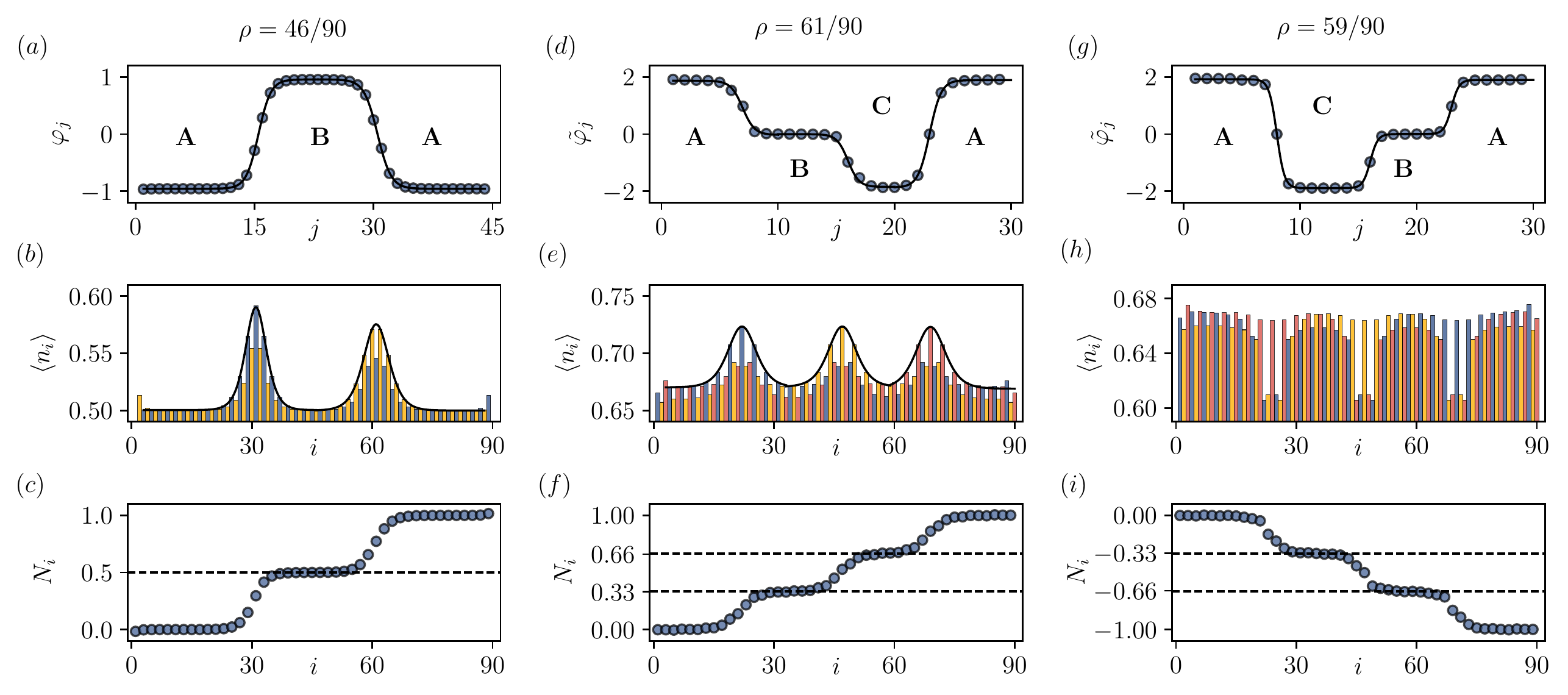}
\caption{\label{fig:fractionalization_long} \textbf{Fractionalization of bound quasi-particles: } {\bf (a)} We represent the order parameter $\varphi_j$~\eqref{eq:Z2_order_parameter} at each unit cell $j$ for the ground state configuration when we add one extra particle on top of half-filling. In this case a soliton-antisoliton pair is created, dividing the chain into ABA  consecutive sectors. In {\bf (b)}, we show the bosonic occupation $\langle n_i \rangle$ in real space. We observe how, inside the bulk of each sector, the occupation corresponds to $0.5$. Around the defects, however, we find peaks where the occupation increases, with a different profile for each sublattice. The black line corresponds to a fit to Eq.~\eqref{eq:Z2_boson_mode}. {\bf (c)} The integrated particle number $N_i = \sum_{j<i} \langle n_j \rangle$ shows how each peak is associated with half a boson. Each  bound bosonic quasi-particle is, thus, fractionalized. The situation is similar around two-third filling when adding or substracting a single particle above  {\bf (d)}, or below   {\bf  (g)} the corresponding filling. We represent the order parameter $\tilde{\varphi}_j$~\eqref{eq:Z4_order_parameter}, showing how three defects appear, creating the SSB patterns ABCA, or ACBA, respectively. The bosonic occupation shows {\bf (e)} peaks and {\bf (h)} drops with respect to $2/3$ around the defects (which agains follows accurately Eq.~\eqref{eq:Z2_boson_mode}). In this case, the each extra boson is fractionalized into three bound quasi-particles with associated particle number {\bf (f)} $ 1/3$ and {\bf (i)} $-1/3$. The calculations were performed using a chain of $L = 90$ sites and taking $U = 10\,t$. In the first column we took $N=46$ particles, $\Delta = 0.80\, t$ and $\beta = 0.02\,t$. In the center and right columns we used $\Delta = 0.85\, t$ and $\beta = 0.01\,t$, with $N = 61$ and $N = 59$ particles, respectively. The solitons were pinned as discussed in the previous section.}
\end{figure*}

So far, we have only focused  on the Ising-spin sector, and proved  the existence of various topological solitons in the groundstate of the $\mathbb{Z}_2$BHM. Although the existence of these solitons is triggered by the coupling of the $\mathbb{Z}_2$ fields to the bosonic quantum matter, we recall that these  solitons can be fully understood classically, as they are characterized by the topological charge of a classical field, i.e. the order parameter of Eq.~\eqref{eq:top_charge}. In this section, we will explore the bosonic sector, and show that there are bound quasi-particles localized within the soliton, which clearly display the bosonic version  of the quantum-mechanical phenomenon of fractionalization. Charge fractionalization was first predicted for relativistic quantum field theories of fermions coupled to  a solitonic background~\cite{relativistic_solitons,NIEMI198699}, and find a remarkable  analogue in the physics of conjugated polymers~\cite{CB_solitons_GN}.  As outlined previously, the direct  experimental observation of this effect would be a real breakthrough in the field, overcoming past limitations in  lightly-doped and disordered polyacetylene~\cite{laughlin_fractional}. 

Arguably, the clearest manifestation of  fractionalization arises by doping a Peierls-type system with a single   particle above/below a given commensurate filling, such as half filling. As it turns out, in order to accommodate for this additional particle,  the groundstate of the $\mathbb{Z}_2$BHM develops a soliton/anti-soliton pair of the $\mathbb{Z}_2$ fields, each of which hosts a bound quasi-particle/quasi-hole with a fractionalized  number of bosons, i.e. the boson splits into two halves. We can, once more,  build our understanding starting from the $\beta=0$ limit, where the soliton/anti-soliton pair becomes two consecutive domain walls. To  analytically investigate the  properties of such domain walls, we have considered the following spin patterns on a chain with even number of sites $L$ and periodic boundary conditions: {\it (i)} the homogeneous N\'eel order, {\it (ii)} a pair of SS domain walls , and {\it (iii)} a pair of  WW domain walls  and, finally, {\it (iv)} one SS and one WW domain wall, all of them separated by a large distance.

In the  $U\rightarrow\infty$ limit, the diagonalization of the model for any Ising spin configuration can be carried out semi-analytically, and gets reduced to a simple matrix diagonalization. This allows us, in particular, to compare the exact groundstate energies  {\it (i)}-{\it (iv)} to understand the nature of the groundstate. Setting $\alpha=0.5t$, we obtain $\Delta_{\rm c}^-=0.368t$ and $\Delta_{\rm c}^+=0.905t$, and set $\Delta$  within this range. We  now dope the system with one extra boson/hole, which can stabilise the configurations {\it (ii)-(iv)} in detriment of {\it (i)} that controls the exact half-filling regime. Comparing the analytical gorundstate energies, we find that  they are degenerate at $\Delta=0.534t$. On the otehr hand, for $\Delta$ closer to $\Delta_{\rm c}^\pm$, the energy differences in energies are very large, and only one of the configurations will be stablished in the groundstate. Specifically, for $\Delta=0.8t$, we find that the configuration {\it (iii)} of two WW domain walls has a considerably lower energy, while for $\Delta=0.4t$ it is configuration {\it (ii)} of two SS domain walls which arises in the groundstate.
 
Interestingly, these results provide some further confirmation of the existence of the Peierls-Nabarro barriers discussed in the preceding section (see Figs.~\ref{fig:loc_length_beta}{\bf (a)-(b)})~\cite{Peierls_nabarro_1,Peierls_nabarro_2}.  By analyzing the semi-analytical energies,  we see that for $\Delta\approx 0.4$($\Delta\approx 0.8$) there is large energy penalty for  finding SS (WW) domain walls in the groundstate, which consists of a pair of WW(SS) domain walls.  This is indeed a manifestation of  some underlying Peierls-Nabarro barrier, since moving the soliton by a single lattice site requires changing its nature from WW(SS) to SS(WW), and this is energetically penalized due to the aforementioned energy differences.  We find that for $\Delta\approx 0.53t$, the Peierls-Nabarro barriers vanish, and soliton hopping occurs through first-order perturbation process proportional to $\beta$, leading to widening of the domain walls forming a soliton/anti-soliton configuration. Conversely, or $\Delta\approx 0.4$($\Delta\approx 0.8$,  the barriers restrict soliton hopping, which can only  occur  virtually as a second-order process proportional to $\beta^2$.

 Let us now contrast these theoretical predictions with numerical results. Figure~\ref{fig:fractionalization_long} (left column) contains all the numerical evidence that supports this mechanism in a chain of $L=90$ sites filled with $N=46$ bosons. Note that this yields precisely an extra particle above half filling $\rho=N/L=1/2+1/L=\rho^\star +1/L$. 

As shown in Fig.~\ref{fig:fractionalization_long}{\bf (a)}, the Ising sector of the $\mathbb{Z}_2$BHM groundstate clearly displays the predicted soliton/anti-soliton profile ABA, and its finite width is a result of the finite quantum fluctuations of the Ising spins. In Fig.~\ref{fig:fractionalization_long}{\bf (b)}, we represent the bosonic particle number for the different sites of the chain. As can be observed, away from the soliton/anti-soliton, the average particle number is consistent with the half-filling condition. Remarkably, as one approaches the topological defects, a build-up in density becomes apparent, which signals the presence of a quasi-particle bound to the topological soliton/anti-soliton.  Moreover,  the density profile of this fractional quasi-particle can be accurately fitted to 
 \beq
 \label{eq:Z2_boson_mode}
 \langle :n_j:\rangle= \langle n_j\rangle-\rho^\star=\frac{1}{4\xi}\sech^2 \left(\frac{j-j_{\rm p}}{\xi}\right),
 \eeq 
where $j = 2i$ or $j=2i+1$ is the sub-lattice index, and $\rho^\star$ is the closest commensurate filling (i.e. $\rho^\star=1/2$ in this case). Remarkably, this expression coincides exactly  with the profile of the zero-modes for the relativistic quantum field theory of fermions~\cite{CB_solitons_GN}, providing a clear instance of universality: regardless of having bosonic/fermionic matter coupled to a $\mathbb{Z}_2$/scalar field, the profiles of the  solitons and the fractionalized quasi-particles  are completely equivalent.

Finally, in order to display clearly the fractionalization phenomenon, let us compute   the integrated number of bosons above the half-filled vacuum
\beq
\label{eq:int_density}
N_i=\sum_{j\leq i} \langle :n_j:\rangle, 
\eeq
 As shown in Fig.~\ref{fig:fractionalization_long}{\bf (c)}, this integrated boson number displays two clear plateaux connected by jumps of $1/2$ in the density. Accordingly,  the localized  quasi-particles of Fig.~\ref{fig:fractionalization_long}{\bf (b)} indeed carry a fractional number of bosons, namely $1/2$ boson each, that is spread within the soliton according to Eq.~\eqref{eq:Z2_boson_mode}.

A similar situation occurs when we dope with extra particles or holes around two-third filling, as it can be appreciated in the center and right columns of Fig.~\ref{fig:fractionalization_long}. In both cases, three defects appear in the ground state (see Fig.~\ref{fig:fractionalization_long}{\bf (d)} and {\bf (g)}), separating different SSB sectors with a density of $2/3$ far from the defects. The latter are associated with peaks or drops in the density that are localized within the different solitons/anti-solitons (see Figs.~\ref{fig:fractionalization_long}{\bf (e)} and {\bf (h)}). Moreover, the integrated density of bosons~\eqref{eq:int_density} displayed in Figs.~\ref{fig:fractionalization_long}{\bf (f)} and {\bf (i)} is fully consistent with the fractional $\pm 1/3$ bosonic quasi-particles.

\begin{figure}[t]
  \centering
  \includegraphics[width=0.9\linewidth]{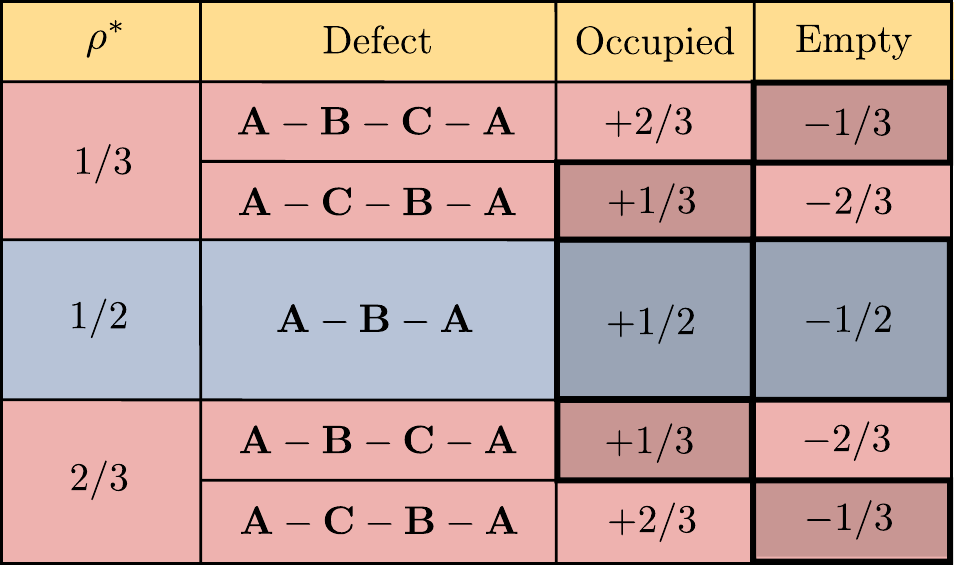}
\caption{\label{fig:fractionalization} \textbf{Topological defects and fractionalized states: } The table  summarizes the particle numbers associated to the localized states bound to the topological solitons, when the state is occupied (particle) or empty (hole). We write defect A-B-C-A, for example, referring to the defects corresponding to each pair of consecutive  sectors: AB, BC and CA, which have the same particle numbers. We highlight the configurations that are generated spontaneously by doping the commensurate densities $\rho^\star$ with one particle or hole. In shaded areas, we highlight the instances   realised in the groundstate.}
\end{figure}

In Figure~\ref{fig:fractionalization}, we summarize the different types of topological solitons and the fractional particle number of the associated quasi-particles. Around two-third filling, in particular, the defects can host fractional bosons with particle number $\pm 1/3$ and $\pm 2/3$. In addition to the situation discussed in the previous paragraph, we note that one extra particle could also be fractionalized into two $+2/3$ and one $-1/3$ quasi-particles. However,   this configuration would not be homogeneous, and is not spontaneously realised in the groundstate, where we have only found  configurations associated to $\pm 1/3$  fractionalized quasi-particles. We note, however, that these configurations can still occur as excitations, and are relevant to understand the possible soliton scattering discussed in the previous section, which must be consistent with the fractionalization phenomena. Recall that the AB and BC defects can collide and lead to a larger conserved charge $Q_{\rm AB}+Q_{\rm BC}=-2$, which corresponds to a new type of anti-soliton AC with charge $Q_{\rm AC}=-2$. This picture is also consistent with the bound  number of bosons, as the AC defect of Fig.~\ref{fig:fractionalization} is associated to a fractional value of $2/3 = 1/3 + 1/3$.

\subsection{Polaron excitations and fractional soliton lattices}

Let us now comment on the possibility that the extra bosons/holes about the commensurate fillings, instead of fractionalizing into  quasi-particles bound to the solitons/anti-solitons, lead  to a  simpler excitation: a topologically-trivial  polaron. In fact, in the fermionic SSH model, a single fermion above the half-filled groundstate does not lead to the fractionalized  pair of quasi-particles. Instead, a lower-energy quasi-particle is formed,  corresponding to the electron being surrounded by a cloud of phonons, the so-called electronic polaron~\cite{SS_soliton_dynamics,CB_solitons_GN}. We note that, in the context of the SSH model, this polaron solution can be understood as a confined soliton-antisoliton pair. The   separation between both defects $d$  is smaller than their corresponding  widths $\xi$, such that the fermion is not fractionalized into a pair of 1/2 charges  bound to each defect, but  instead distributed across the entire polaron with the same total charge~\cite{CB_solitons_GN,ssh_review}. If one studies the energy of the soliton-antisoliton pair as a function of the distance $E_{\rm  g}(d)$, a global energy minimum is found for the separation of the polaron mentioned above~\cite{KIVELSON1986301}.

 \begin{figure}[t]
  \centering
  \includegraphics[width=1\linewidth]{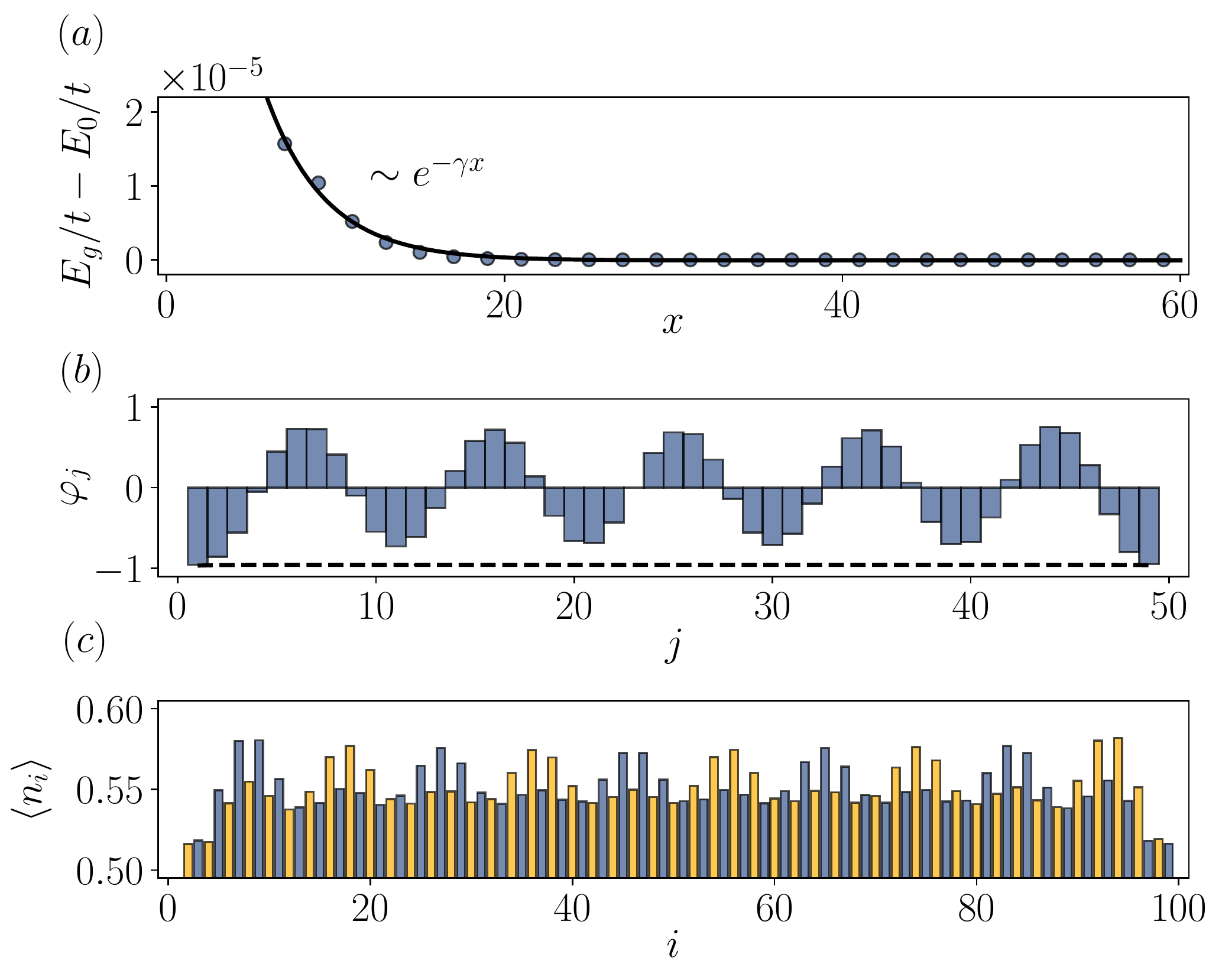}
\caption{\label{fig:soliton_lattice} \textbf{Polarons and the soliton lattice}: {\bf (a)} Groundstate energy $E_{\rm g}(x)$ for a chain of $L=120$ sites filled with $N=62$ bosons, as a function of the distance $x$ between the pinned soliton-antisoliton pair. The absence of a miniminum for a given distnace $x_0$ shows that topologically-trivial polarons are absent in the groundstate of the $\mathbb{Z}_2$BHM.  {\bf (b)} The order parameter $\varphi_j$ \eqref{eq:Z2_order_parameter} for a chain of $L=100$ sites filled with $N=55$ bosons self-assembled in a periodic configuration of soliton-antisolitons, maximizing the inter-soliton distance, and leading to the so-called soliton lattice. The dotted line represents the corresponding staggered polarization of the half-filled groundstate without any topological defect.  {\bf (c)} The local boson density displays the localized nature of the bound quasi-particles, which again display a fractionalized density upon a closer inspection.}
\end{figure} 

 In the $\mathbb{Z}_2$BHM, we  can calculate numerically the groundstate energy as a function of the soliton-antisoliton distance $E_{\rm  g}(d)$, which is controlled though the pinning centers of the perturbation~\eqref{eq:pinning_term}. In Fig.\ref{fig:soliton_lattice}{\bf (a)}, we show how this energy does not present any global minimum corresponding to a polaronic solution, which in this case would stand for the bosonic particle surrounded by spin-wave fluctuations. 
 
 According to this result, we can rule out the existence of any confining mechanism, and ensure that the groundstate corresponds to the distant soliton-antisoliton pair with fractionalized bound quasi-particles. This results are also useful to discuss the following type of groundstate, the so-called soliton lattice. Since the energy decreases exponentially with the soliton distance (see the fit in Fig.\ref{fig:soliton_lattice}{\bf (a)}), solitons tend to repel each other seeking for groundstate configurations with the maximal inter-soliton distance. If we keep on adding additional particles above/below half-filling, but do not fix their relative positions by the external pinning~\eqref{eq:pinning_term}, the solitons-antisolitons pairs will self-assemble in a crystalline configuration that maximises the inter-soliton distance: a soliton lattice. Such type of solutions were originally predicted  for the SSH model of polyacetylene~\cite{PhysRevLett.46.742, PhysRevB.35.734}, and are also known as kink-antikink crystals  in relativistic field theories of self-interacting fermions at finite densities~\cite{PhysRevD.67.125015, PhysRevLett.100.200404}.

  \begin{figure*}[t]
  \centering
  \includegraphics[width=1.0\linewidth]{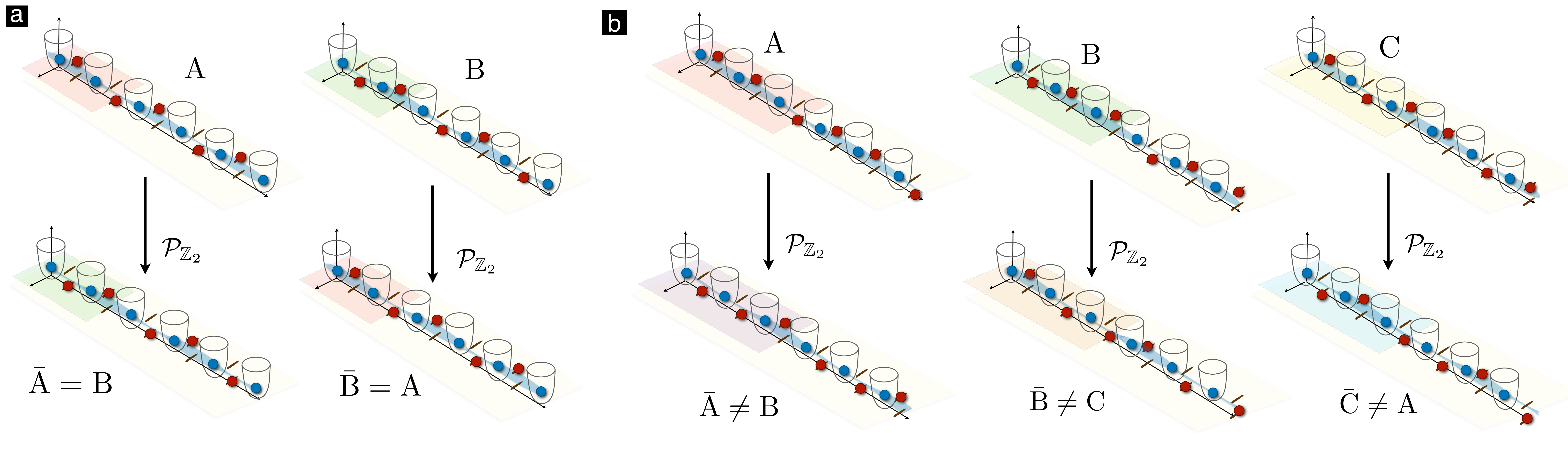}
\caption{\label{fig:tbow} \textbf{Topological Bond Order Wave phases: } {\bf (a)} The Bond Order Wave phase at half filling (BOW$_{1/2}$) is doubly degenerate. The two SSB sectors, A and B, are related by a one-site translation, and can be distinguished by their topological properties: while A is topologically trivial, B is a symmetry-protected topological phase \cite{z2_bhm_2}. Note that, by applying a $\mathcal{P}_{\mathbb{Z}_2}$ inversion  $\sigma^z \rightarrow -\sigma^z$, we can transform states in the A sector to states in the B sector and viceversa. {\bf (b)} This property is not satisfied, however, at one-third and two-third fillings. For each of these densities, we find again a BOW phase with a three-fold degeneracy. A, B and C denote the three SSB sectors, connected again by one-site translations. Note that now the $\mathcal{P}_{\mathbb{Z}_2}$ transformation takes states from these sectors to a different phase. We denote the later as TBOW$_{1/3}$ (respectively TBOW$_{2/3}$), which is again three-fold degenerate  $\bar{\rm A}$, $\bar{\rm B}$ and $\bar{\rm C}$, and appears at stronger Hubbard interactions~\cite{z2_bhm_3}.  The latter is a symmetry-protected topological phase, as opposed to the BOW$_{1/3}$ (respectively BOW$_{2/3}$), which possesses a zero topological invariant~\cite{z2_bhm_3}. The fact that, as opposed to the half-filled case, $\mathcal{P}_{\mathbb{Z}_2}$ do not coincide  with the one-site translation has  an important consequence: while in the former case topological defects separate regions with different topological properties in the bulk, this is not true for the latter, and localized states associated to the defects are not expected a priori based on topological arguments. We will see, however, that a topological origin for such states can be recovered by extending the system to two dimensions through a pumping mechanism.}
\end{figure*}

 In Fig.~\ref{fig:soliton_lattice} {\bf (b)}, we represent the Ising sector of the $\mathbb{Z}_2$BHM groundstate for a chain with $L=100$ sites and $N=55$ bosons. As can be readily appreciated, in the absence of pinning~\eqref{eq:pinning_term}, the $\mathbb{Z}_2$ fields self-assemble in a periodic configuration of soliton-antisoliton pairs that delocalize over the chain while maximising the inter-soliton distance. In this way, the background spins form a soliton lattice, and lead to a periodic configuration of fractionalized bosonic  quasi-particles: a fractional soliton lattice [see Fig.~\ref{fig:soliton_lattice}{\bf (c)}]. As can be observed  in Fig.~\ref{fig:soliton_lattice} {\bf (b)}, the value of order parameter in between a consecutive soliton-antisoliton pair does not reach the value of the defect-free configuration at precisely half-filling. As occurs for the SSH model~\cite{PhysRevLett.46.742, PhysRevB.35.734}, this can be considered as evidence that the energy gap of the soliton lattice is smaller than that of the defect-free configuration or the pinned-soliton groundstate, signalling that the topological defects in the soliton lattice are not independent excitations above the perfectly-dimerised groundstate. Instead, the topological defects in the soliton lattice are coupled and form an energy band, leading to groundstates that are fundamentally different from a collection of uncoupled topological solitons.  We would like to emphasize that, although this type of solutions has been analytically predicted before~\cite{PhysRevLett.46.742, PhysRevB.35.734,PhysRevD.67.125015, PhysRevLett.100.200404}, our DMRG results for the $\mathbb{Z}_2$BHM provide, to the best of our knowledge, the first numerically-exact confirmation of their existence without the approximations underlying previous analytical works (e.g. large-$N$ methods) .

\section{\bf Bulk-defect correspondence}
\label{sec:bulk-defect}

So far, we have solely focused on the topological properties of the $\mathbb{Z}_2$-field sector, and discussed the fractionalized nature of the bound bosonic quasi-particles. In this section, we delve into  the full topological  characterization of the defects, and show how the fractionalized matter can bring an extra topological protection to the groundstate, making connections to the notion of symmetry-protected topological defects (SPT-d). We remark, once more,  that previous studies typically consider the properties of SPT-d in the presence of externally-adjusted static solitons~\cite{topological_defects_1, topological_defects_2}. Here, on the other hand, we focus on the full many-body problem where the solitons have their own dynamics, and the back action of the matter sector on the solitons becomes relevant. Moreover, we will unveil the particular topological origin of the associated fractional bosons  through a generalized bulk-defect correspondence.

\subsection{ Symmetry protection of  bound  quasi-particles}

Paralleling our discussion of the previous sections, let us start with the simplest situation: the  half-filled configuration, where the Peierls' mechanism gives rise to a two-fold degenerate groundstate with a dimerization of the $\mathbb{Z}_2$ field (see Fig.~\ref{fig:tbow}). In the hardcore limit $U/t\rightarrow\infty$, the bosons can be mapped onto fermions by means of a Jordan-Wigner transformation~\cite{Jordan1928}, which shows that the resulting groundstate has  both inversion symmetry and a sub-lattice, so-called chiral, symmetry. According to the general classification of symmtery-protected topological phases~\cite{classification_spt}, the bulk TBOW$_{1/2}$ is a clear instance  of a \textsc{BDI} topological insulator~\cite{z2_bhm_2}. The two degenerate ground states A and B can be characterized as trivial or topological, respectively, with the help of a topological invariant, namely the Berry phase $\gamma$~\cite{doi:10.1098/rspa.1984.0023}. The latter takes the value $\gamma=0$ in the trivial  A phase, and $\gamma=\pi$ in the topological  B phase. Furthermore, the chiral symmetry ensures the so-called bulk-edge correspondence: a groundstate that displays a Berry phase $\gamma=\pi$ will have one edge state at each boundary of the chain.

\begin{figure}[t]
  \centering
  \includegraphics[width=1.0\linewidth]{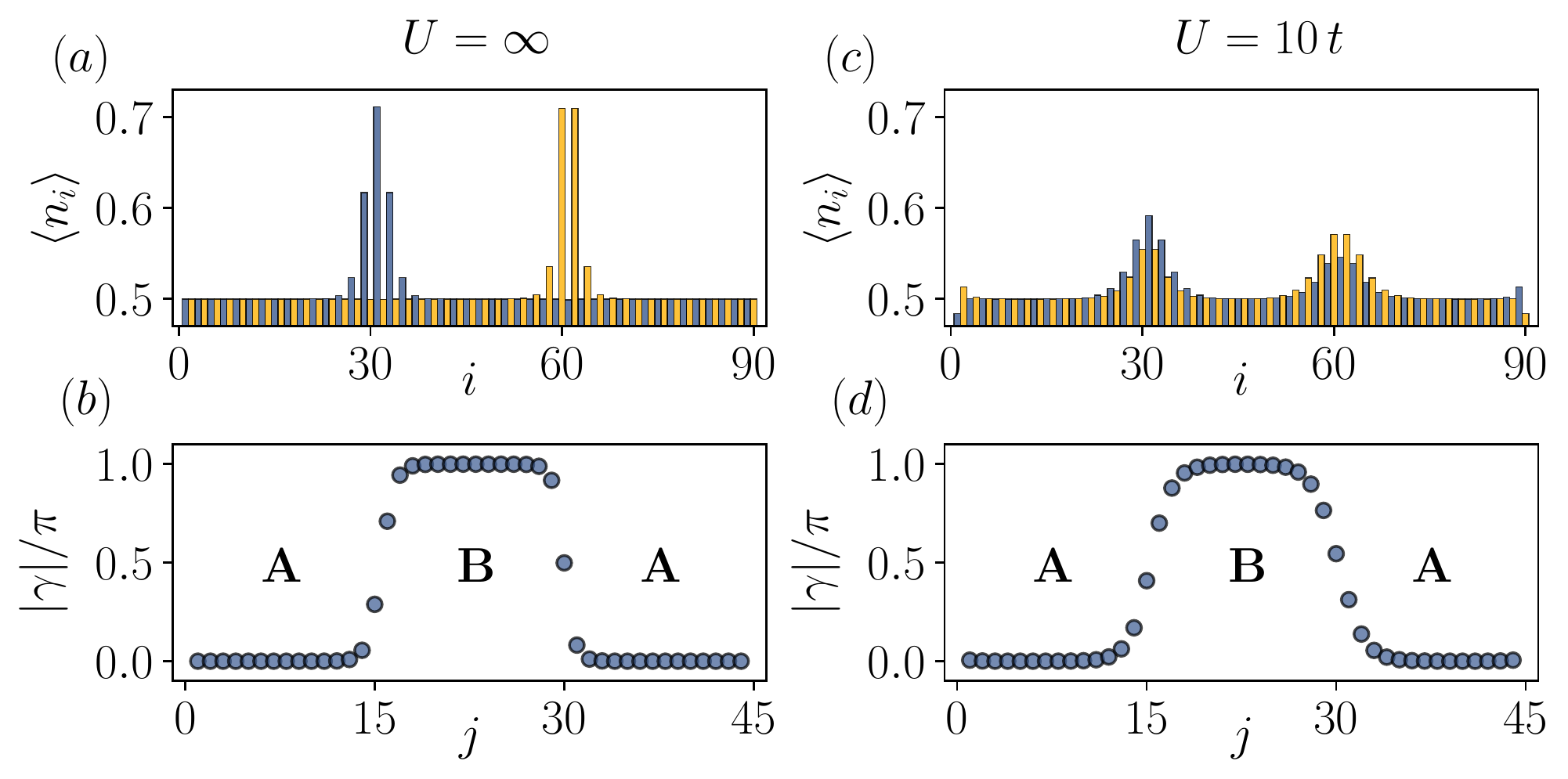}
\caption{\label{fig:spt_dimer} \textbf{Topological invariant and fractionalization: } {\bf (a)} Occupation number $\langle n_i \rangle$ for a chain with $L = 90$ sites in the hardcore limit, $U/t \rightarrow \infty$, when we dope a state in the BOW phase with one extra particle above half filling. At the location of the defects, two peaks in the occupation can be observed, each one localized only in one sublattice (represented in different colors) as a consequence of chiral symmetry. {\bf (b)} Local Berry phase $\gamma$ calculated on the bonds that separate different unit cells $j$. This quantity is quantized to $0$ and $\pi$ in the different SSB sectors, and interpolates between these two values in the region where the defects are located. {\bf (c)} For a finite value of $U$ we can still observe peaks in the occupation. These are no longer localized in specific sublattices, since chiral symmetry is broken. However, the topological Berry phase is still quantized far from the defects {\bf (d)} since inversion symmetry is still preserved. In the hardcore limit we used the parameters $\Delta = 0.70t$ and $\beta = 0.03t$, and for the softcore case $\Delta = 0.80t$ and $\beta = 0.02t$.}
\end{figure}

Let us now revisit the  case of one extra particle above half filling. Figure~\ref{fig:spt_dimer}{\bf (a)} shows the real-space  occupation of bosonic in the ground state of a finite system of $L = 90$ sites. The system displays the pattern A-B-A with two solitons separating the degenerate SSB configurations. To understand the topological origin of these states, we show in Fig.~\ref{fig:spt_dimer}{\bf (b)} our numerical calculation of the local Berry phase in real space~\cite{hatsugai_2006},   computed on the intercell bond (i.e. the bond joining two neighboring unit cells). As can be observed in this figure, the local Berry phase in the configuration A is equal to $\gamma=0$, while the one in the  B configuration yields $\gamma=\pi$. In such a situation, the theory of topological defects~\cite{topological_defects_1, topological_defects_2} predicts that   localized and topologically-protected boundary states will appear at the interface of the two topologically-distinct regions. Furthermore, as a consequence of chiral symmetry, these states have support  in just one of the two sub-lattices. This makes them robust against perturbations that respect the chiral symmetry, but are not sufficiently strong to close the gap of the system. Therefore, apart from the inherent robustness of the classical topological solitons, the total defects formed by a soliton and a fractionalized  bosonic quasi-particle are also protected against chiral-preserving perturbations, leading to the aforementioned SPT-d~\cite{topological_defects_1, topological_defects_2}. 

From a pragmatic point of view, these SPT-d constitute an alternative to observe topological edge states in a cold atom experiment. We note that the presence of topological edge states at the boundaries of a cold-atom system can be sometimes hampered by the presence of an additional trapping potential~\cite{PhysRevA.82.013608}. Previous attempts to overcome these difficulties rely on externally adjusting inhomogeneous configurations~\cite{PhysRevLett.105.255302}, imposing background solitonic profiles on a superlattice structure~\cite{grusdt_2013}, or by shaking the optical lattice~\cite{Przysina2015}. Here, on the contrary, the topological solitons  are dynamically generated by doping the system, and self-adjust to certain positions of the chain depending on the  doping. In particular, there are certain configuration where the SPT-d can be found at the middle of the system (see Fig.~\ref{fig:loc_length}), where the deleterious effect of the  trapping potential would be absent.

Having clarified the topological protection of these SPT-d in the simpler half-filled  and hardcore limits, let us now turn into more complex situations, and assess the nature of the topological protection of these defects {\it (i)} away from the hardcore constraint, or {\it (ii)} around other fractional fillings.

For finite values of the Hubbard repulsion $U$, the chiral symmetry is explicitly broken even in the  half-filled case. Nevertheless, the system still possesses inversion symmetry, which can lead to a quantized non-zero Berry phase and an intertwined topological phase. We emphasize that the bulk-boundary correspondence is no longer guaranteed for these phases:  the edge states break the inversion symmetry, and are therefore not protected by the topology of the bulk~\cite{z2_bhm_2}. Let us now discuss the situation for fillings above/below half-filling. Figure~\ref{fig:spt_dimer}{\bf (c)} shows the real-space bosonic occupation for $U=10t$. The soliton-antisoliton pair is still present for such finite interactions but, as a direct consequence of the loss of  chiral symmetry, the support of the fractionalized modes is no longer  restricted to a single sub-lattice. Nonetheless, Figure~\ref{fig:spt_dimer}{\bf (d)} shows that the local Berry phase  is still quantized to  $\gamma=0$ and $\gamma=\pi$ in the bulk of the trivial and topological configurations, respectively. This quantization is preserved even in the absence of chiral symmetry,  as there still exists the discrete  inversion symmetry. We observe how, in the region where the solitons interpolate between the two inversion-symmetric groundstates,  and  inversion symmetry is thus not maintained, the Berry phase attains intermediate values  connecting  the two quantized values that appear far away from the soliton cores. In summary, the bulk can be characterized by a topological invariant, but there is no direct bulk-defect correspondence responsible for the  protection of the defects. In the following section, we shall revisit this scenario in search for a generalised bulk-defect correspondence.

Let us now turn to two-third filling, where we recall that the Peierls' mechanism gives rise to a threefold degenerate groundstate with a trimerization of the $\mathbb{Z}_2$ fields (see Fig.~\ref{fig:tbow}). The resulting phases are insulators that can either be topologically trivial or non-trivial, depending on the strength of the Hubbard interactions~\cite{z2_bhm_3}. Figure~\ref{fig:tbow}{\bf (b)} depicts the different ground state configurations for $U=10t$ (trivial) and $U=15t$ (topological): the trivial phase has a BOW pattern with two strong bonds and one weak bond (A), whereas the topological phase has a BOW pattern with two weak bonds and one strong bond ($\bar{\rm A}$). The topology of these ground states can be characterized with the help of the inter-cell local  Berry phase, which is equal to $\gamma=0$ in the trivial phases A, B and C, but non-zero $\gamma=\pi$ in the  topological phases $\bar{\rm A}$, $\bar{\rm B}$ and $\bar{\rm C}$. We note that the three degenerate ground states A, B and C (resp.  $\bar{\rm A}$, $\bar{\rm B}$ and $\bar{\rm A}$) are related to each other by a translation of one site, and are thus equivalent in the thermodynamic limit. We emphasize here that the three ground states have the same topology, unlike the half-filled case, where B = $\bar{\rm A}$ has a different topology with respect to A (see Fig.~\ref{fig:tbow}). 

We now move away from the commensurate fillings, and address the topological characterization of the bound quasi-particles that appear when doping the system with  one particle above the two-third filled groundstate, which would correspond to  the trivial BOW phase for $U = 10t$. As discussed in previous sections, three topological solitons arise leading to the A-B-C-A configuration [See Fig.~\ref{fig:tbow}{\bf (b)}], each of which hosts a  localized bosonic quasi-particle with a fractional charge of $1/3$. Notice that,  in contrast to  half-filling,  the defects are no longer separating regions with different topological bulk properties. According to the theory of SPT-d, since the solitons do not interpolate between topologically-distinct regions, there is no reason to expect that a  quasi-particle will be bound to the soliton. Nonetheless, such bound quasi-particles do appear, carry fractionalized charges, and we would like to understand if they have some generalised topological origin. In the next section, we show  that these quasi-particles can be understood  as remnants of  topologically protected defect modes of an extended 2D system, even if this cannot be inferred a priori in the 1D system.

\subsection{Quantized inter-soliton pumping}

 In this section, we show how a generalized bulk-defect correspondence can be stablished by extending the solitons to topological defects in a higher dimension through a Thouless pumping argument~\cite{pumping_thouless}.

\subsubsection{Thouless pumping and the bulk-defect correspondence}

\begin{figure*}[t]
\centering
	\begin{subfigure}{}
	\centering
	\hspace{0.1cm}
	\includegraphics[width=0.258\linewidth]{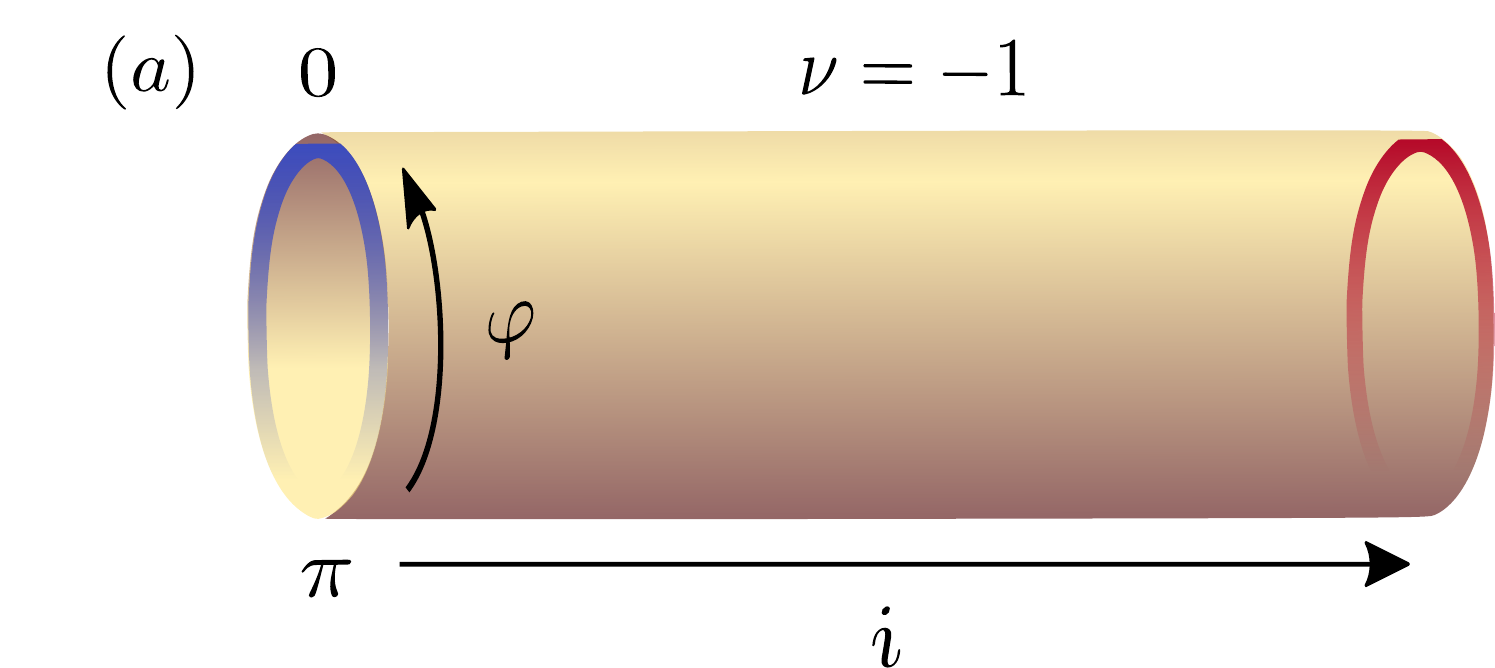}
	\end{subfigure}\hspace*{\fill}
	\begin{subfigure}{}
	\centering
	\hspace{0.6cm}
	\includegraphics[width=0.278\linewidth]{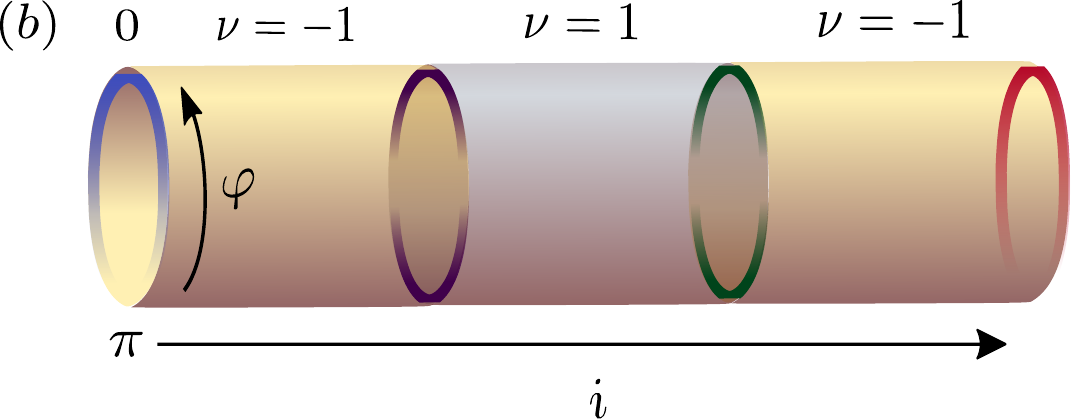}
	\end{subfigure}\hspace*{\fill}
	\begin{subfigure}{}
	\centering
	\includegraphics[width=0.32\linewidth]{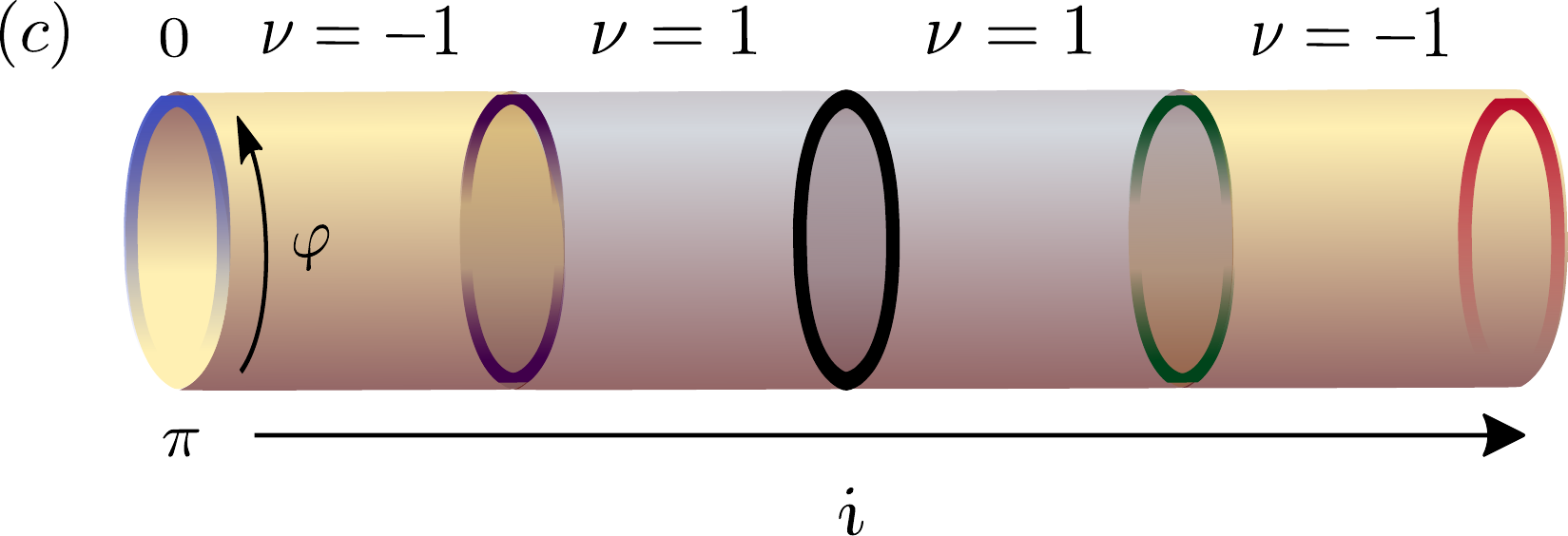}
	\end{subfigure}\hspace*{\fill}
	\medskip
	\begin{subfigure}{}
	\centering
	\includegraphics[width=0.3\linewidth]{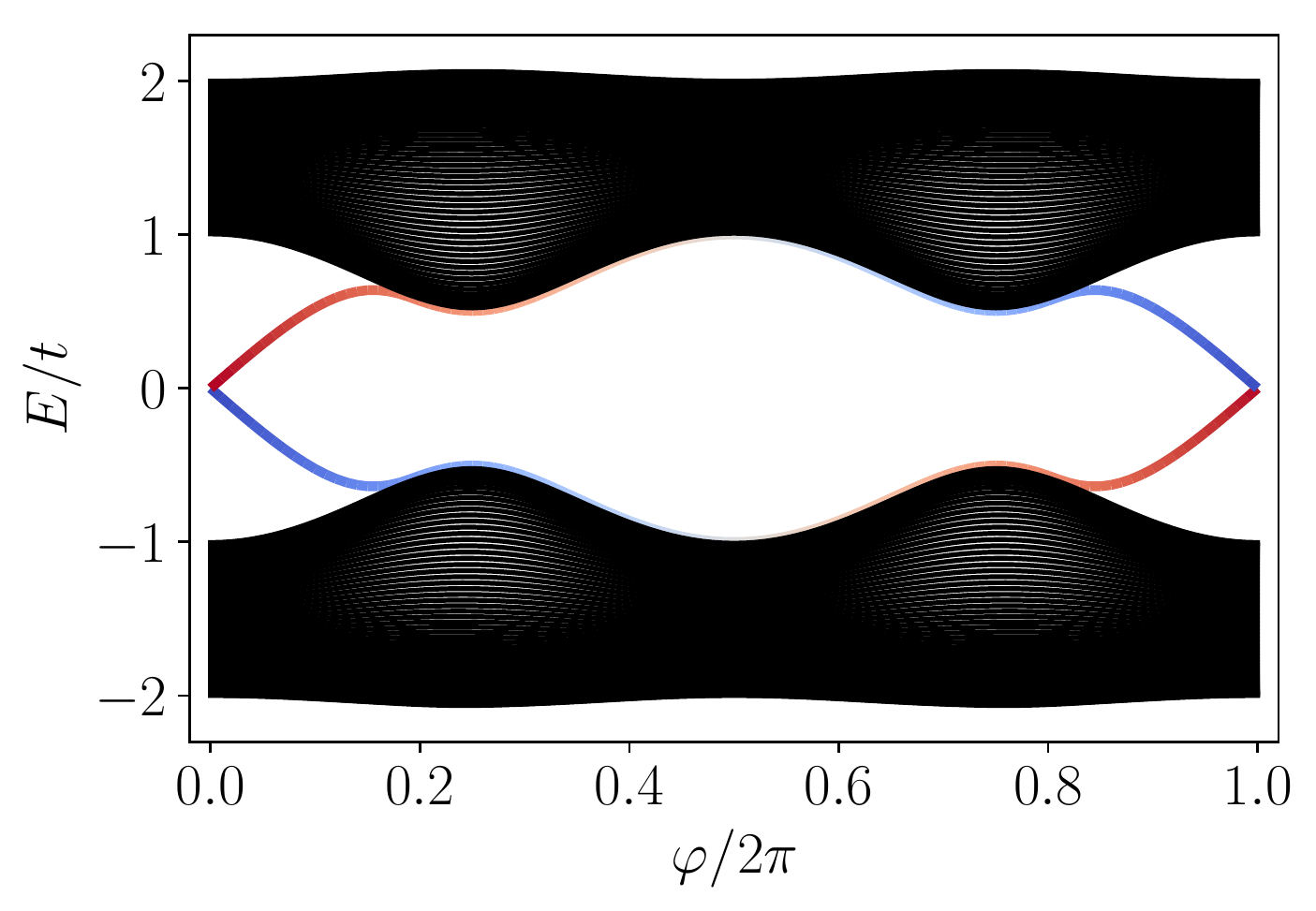}
	\end{subfigure}\hspace*{\fill}
	\begin{subfigure}{}
	\centering
	\includegraphics[width=0.3\linewidth]{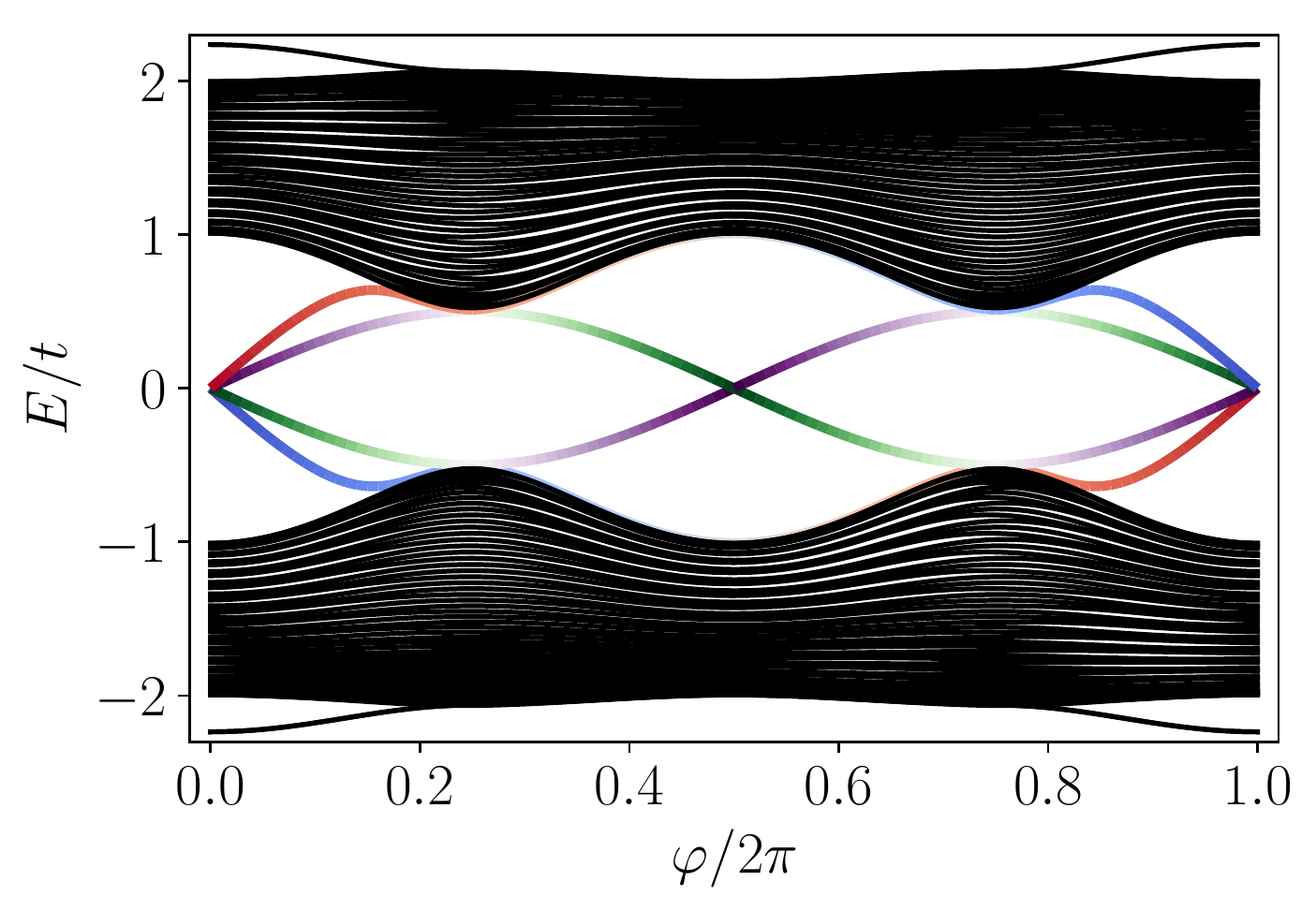}
	\end{subfigure}\hspace*{\fill}
	\begin{subfigure}{}
	\centering
	\includegraphics[width=0.3\linewidth]{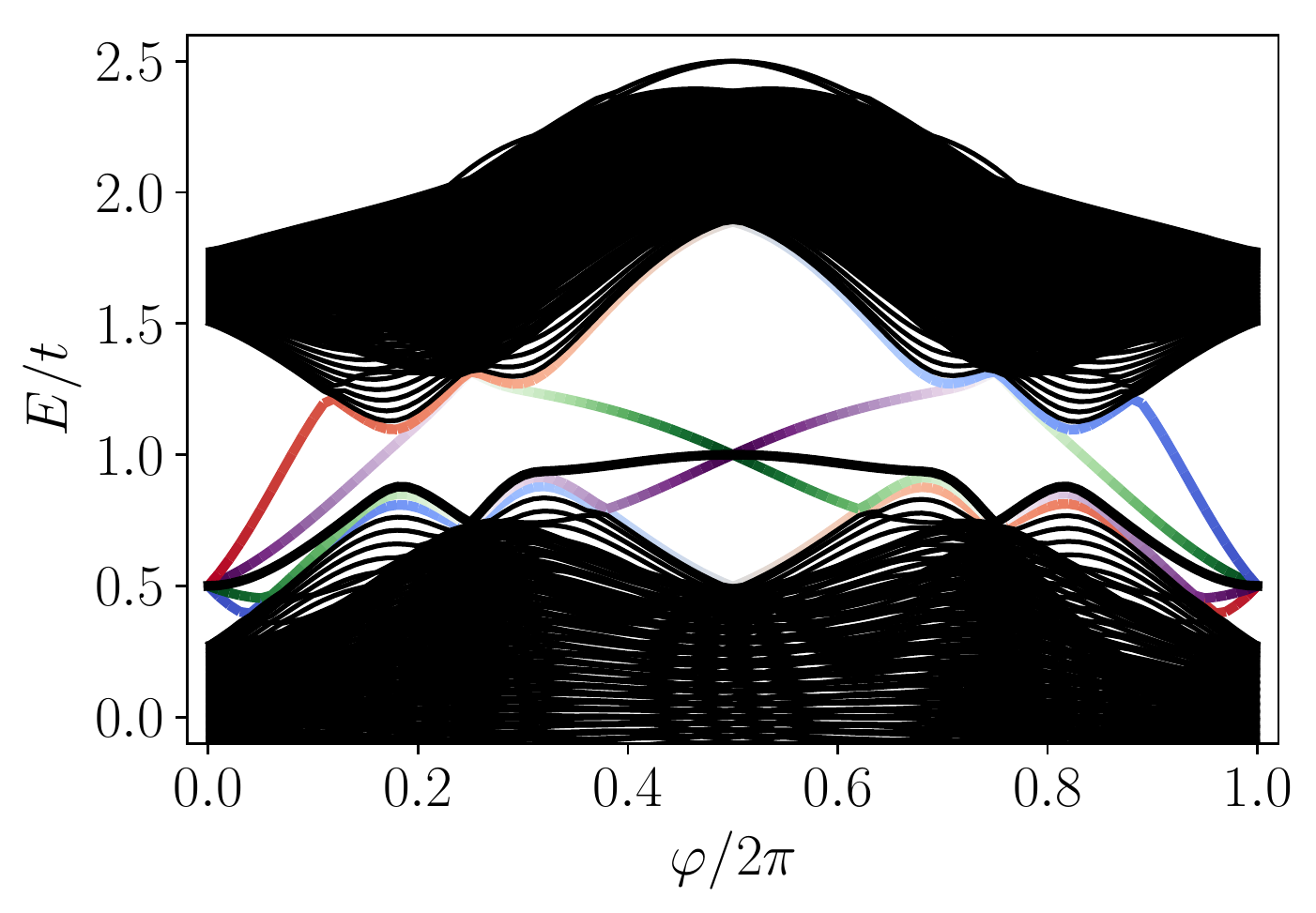}
	\end{subfigure}\hspace*{\fill}

\caption{\label{fig:pumping_dimer}\textbf{Non-interacting pumping in a static lattice: } In the upper panels we show the 2D extension to a finite cylinder of a one-dimensional fermionic chain in a static lattice through a pumping process (eq.~\eqref{eq:H_dimer_pumping}), where the pumping parameter $\varphi$ corresponds to the momentum in the periodic dimension. The associated Chern number in each region is denoted by $\nu$. The lower panels show the corresponding spectral flow along the pumping. Localized edge states inside the gap are represented in different colors that correspond to their position in real space, where the intensity is associated qualitatively to their localization length in 1D. {\bf (a)} Dimerized lattice without defects. Two edge states, each one localized at   one boundary of the system,  cross the gap connecting the two bands. These edge states are topologically protected in the extended 2D system since they can not be removed unless the gap closes. They have the same energy at the inversion-symmetric point $\varphi = 0$, where the system is in the topological configuration $\bar{\rm A}$. The other inversion-symmetric point at $\varphi = \pi$ corresponds to the trivial case A. Along the pumping cycle, one particle is transported through the bulk from one edge to the opposite one, where the direction is different for each state and depends on the Chern number of the extended 2D cylinder. {\bf (b)} Dimerized lattice with two domain walls. Apart from the edge states discussed in the previous case, two extra topologically-protected localized states, each one associated to one domain wall, appear in the gap connecting the bands. Here the two inversion-symmetric points correspond to the configurations $\bar{\rm A}$A$\bar{\rm A}$ and A$\bar{\rm A}$A, respectively. In this case, two particles are pumped from one defect to the other during the cycle, since the difference between the Chern numbers of the regions they separate is now $\pm 2$. {\bf (c)} Trimerized lattice with three domain walls, where we represent the spectrum around the gap that opens at two-third filling. In this case, apart from the two edge states, three extra localized states associated to the defects appear in each gap. However, only two of them connect different bands. The third one (black line, localized at the center of the cylinder) starts and ends in the  valence band, so it is not topologically protected as it can be removed without closing the gap. This can be understood using the bulk-defect correspondence, since that particular defect separates regions with the same Chern number. In this case, the two inversion-symmetric points correspond to the configurations $\bar{\rm A}\bar{\rm B}\bar{\rm C}$ and ABC, respectively.}
\end{figure*}

In the beginning of the 80s, Thouless showed that   gapped 1D systems  undergoing a slow cyclic modulation can display   quantised  particle transport, and that the robustness of this effect is rooted in  an underlying non-zero topological invariant in higher dimensions~\cite{pumping_thouless}. We now briefly review this concept for the so-called Rice-Mele model~\cite{PhysRevLett.49.1455}, a generalisation of the SSH  fermionic model in a static solitonic backround~\cite{berry_phase_review}. We consider the dimerized Hamiltonian with an extra chemical potential term following an adiabatic cycle,
\begin{equation}
\label{eq:H_dimer_pumping}
\begin{aligned}
H(\varphi) =\sum_i - (t + (-1)^i\delta_\varphi)\left(c^\dagger_i c_{i+1} + \text{h.c.}\right)+ (-1)^i \Delta_\varphi \, c^\dagger_i c_i
\end{aligned}
\end{equation}
where $\delta_\varphi = \delta\,\text{cos} (\varphi)$ and $\Delta_\varphi = \delta\,\text{sin} (\varphi)$. Varying the parameter $\varphi$ in time, the groundstate starts in the  topological configuration $\bar{\rm A}$ ($\varphi=0$) and adiabatically evolves into the trivial configuration A ($\varphi=\pi$). The onsite staggering potential is chosen such that the model remains  gapped  during the whole adiabatic cycle. As a consequence of this cyclic evolution, the charge that is pumped across the chain, $\Delta n$, is quantized to integer values. This can be shown by relating this quantity to the  Chern number $\nu$ of an extended 2D system, where the time-dependent parameter $\varphi$ is taken as  the momentum corresponding to an additional synthetic dimension~\cite{berry_phase_review}, namely $\Delta n = -\nu=- \sum_i  \nu_i$, where the $\nu_i = \frac{1}{2\pi} \int_0^T  \, \int_\mathrm{BZ} \, \Omega_i \mathrm{d}\tau \mathrm{d}q$ are the Chern numbers of the occupied bands, written in terms of the Berry curvature $\Omega_i = i(\braket{\partial_q u_i \vert \partial_\varphi u_i} - \braket{\partial_\varphi u_i \vert \partial_q u_i})$ of the corresponding  Bloch eigenstates $\ket{u_n}$. 

Figure~\ref{fig:pumping_dimer}{\bf (a)} shows the full energy spectrum of the single-particle Hamiltonian~\eqref{eq:H_dimer_pumping} in terms of $\varphi$. At $\varphi = 0$ the lattice is in the topological configuration $\bar{\rm A}$, and we find two degenerate edge states in the middle of the gap. Following the bulk-edge correspondence argument~\cite{Hatsugai1993}, the number of topologically protected edge states at each boundary is equal to $\vert\Delta n\vert$ and the sign of $\Delta n$ determines the direction of the charge transport. These states connect the two bands as  $\varphi$ is modified, which is commonly referred to as spectral flow, and are responsible for transporting the charge during the pumping process. We find a Chern number of $\nu_{\bar{\rm A}}=-1$ for this pumping sequence. 

Figure~\ref{fig:pumping_dimer}{\bf (a)} can be interpreted as the spectrum of a 2D system in a cylindrical geometry, where $\varphi$ is the momentum in the synthetic dimension. The localized edge states become, in this picture, the 1D conducting edge states at the boundaries of the synthetic cylinder. The phase of this synthetic 2D system corresponds to a Chern insulator and, as opposed to the 1D case,  the bulk-boundary correspondence guarantees the presence of topologically-protected edge states even in the absence of chiral symmetry. Therefore, this 2D extension through the Thouless pumping establishes a generalized bulk-boundary correspondence, where the localized states in 1D can be seen as remnants of protected edge states in 2D. This is true as long as the 1D bulk has a non-zero topological invariant, as it is the case in our bosonic model, where the topology is protected by inversion symmetry. Abusing the notation, we will associate Chern numbers to the 1D phases, keeping in mind that this requires the specification of not only the initial state for the pumping protocol, but also of its direction. In this case, if we reverse the direction of the pumping we would get $\nu_{\bar{\rm A}}=1$. In the following, we will keep the same pumping protocol, specified by Eq.~\eqref{eq:H_dimer_pumping}.

Let us now use this bulk-boundary correspondence in the context of SPT-d and topological solitons. We first notice that the topological solitons in the bulk of the 1D chain can also be understood as extended interfaces  between regions with different Chern numbers in the adiabatic pumping. We now consider the configuration $\bar{\rm A}$A$\bar{\rm A}$ for $\varphi=0$, and its cyclic evolution  to the configuration A$\bar{\rm A}$A for $\varphi=\pi$ . Far from the defects, the bulk of the three regions can be characterized by different Chern numbers: $\nu_{\bar{\rm A}}=-1$, $\nu_{\rm A}=1$ (see Fig.~\ref{fig:pumping_dimer}{\bf (b)}). 

The theory of topological defects predicts that the  periodic particle pumping across one point of the chain will depend on the Chern numbers of the corresponding regions. In particular, the number of bound states  that circulate around each soliton  is equal to the difference of Chern numbers between the regions connected by the soliton. This can be observed in Figure~\ref{fig:pumping_dimer}{\bf (b)}, where we represent the spectral flow of a finite dimerized chain with OBC and two domain walls. Apart from the two localized states associated to the edges of the chain, two extra localized states cross the gap connecting both bands. These are associated with the domain walls, and cross at the two inversion-symmetric points, where they are degenerate. The extended 2D cylinder consists of different regions with different Chern numbers, and the localized states associated to the solitons in 1D can be interpreted as  topologically-protected conducting states that reside at the 1D circular boundaries between these regions. Note that these  states circulate twice as fast compared to the states on the boundaries, since the   Chern number difference is  doubled. 

The explicit connection between the adiabatic pumping in 1D and the  Chern insulators  in 2D (\ref{eq:H_dimer_pumping}) allows us to clarify the topological origin of the localized states bound to the solitons in the case of a non-interacting fermionic system on a static dimerized chain, generalizing the bulk-edge correspondence to a defect-edge correspondence. We will now apply this reasoning to the full interacting system with dynamical solitons.

\subsubsection{Inter-soliton pumping in the $\mathbb{Z}_2$ Bose Hubbard model}

\begin{figure}[t]
  
\includegraphics[width=1.0\linewidth]{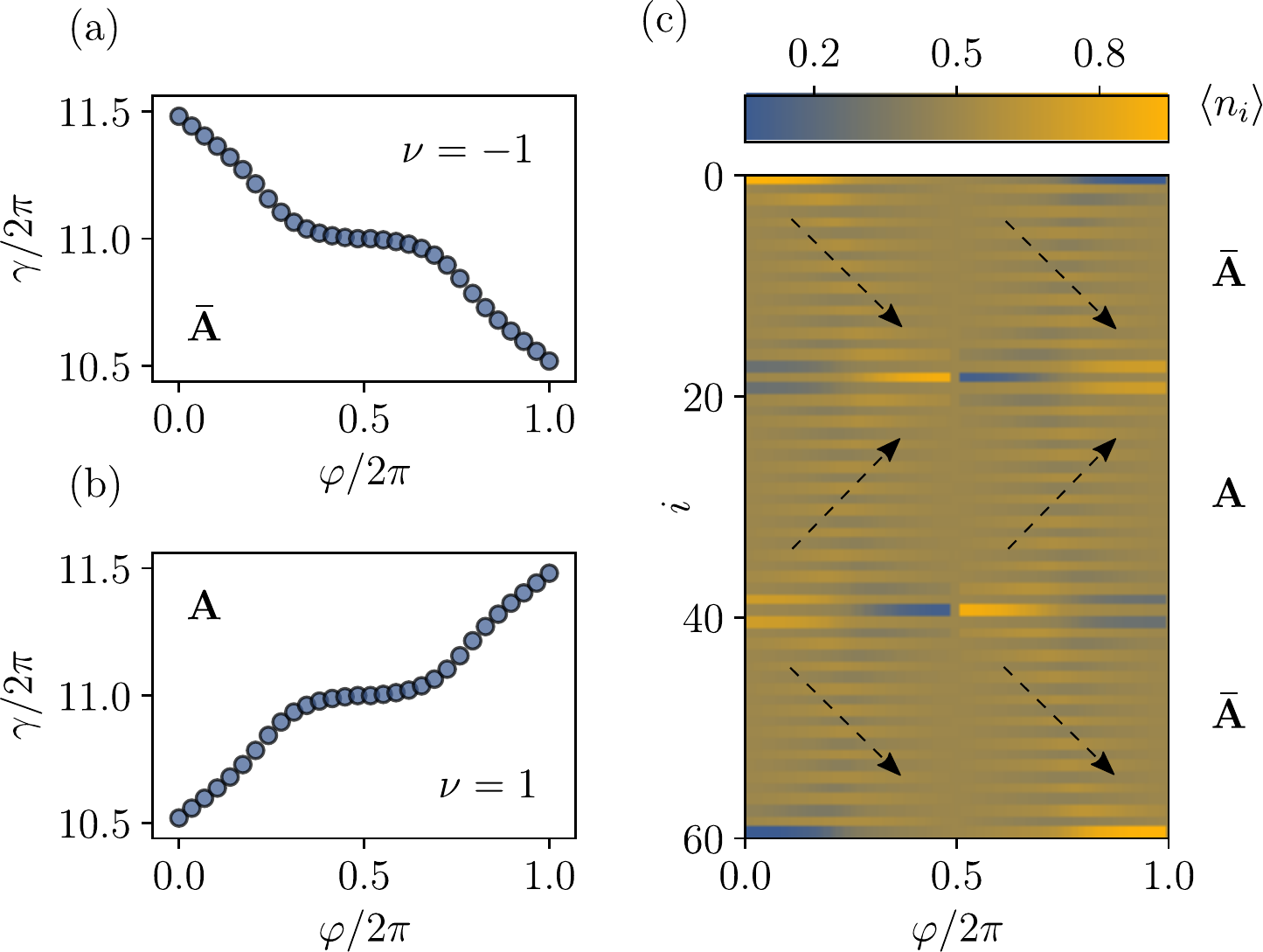}

\caption{\label{fig:chern_entanglement} \textbf{Inter-soliton pumping with $\mathbb{Z}_2$ solitons: } {\bf (a)} and {\bf (b)} show the local Berry phase $\gamma$ calculated at the middle of a finite chain with $L=42$ sites  as a function of the pumping parameter $\varphi$. Both correspond to half filling, but starting from the homogeneous configurations $\bar{\rm A}$ and A, respectively. Using Eq.~\eqref{eq:chern_berry} we can calculate the corresponding Chern numbers, obtaining $\nu_{\bar{\rm A}}=-1$ and $\nu_{\rm A}=+1$, respectively. The sign of the Chern number gives us the direction of the transport, $\Delta n = -\nu$. This can be observed in {\bf (c)}, where we represent the real-space bosonic occupation $\langle n_i \rangle$ through the pumping cycle of a system with two domain walls, starting from the $\bar{\rm A}$A$\bar{\rm A}$ configuration at $\varphi = 0$. Each region transports a quantized charge in the bulk, but the direction (dashed arrows) is different in each of them.}
\end{figure}

\begin{figure}[t]
  
\includegraphics[width=1.0\linewidth]{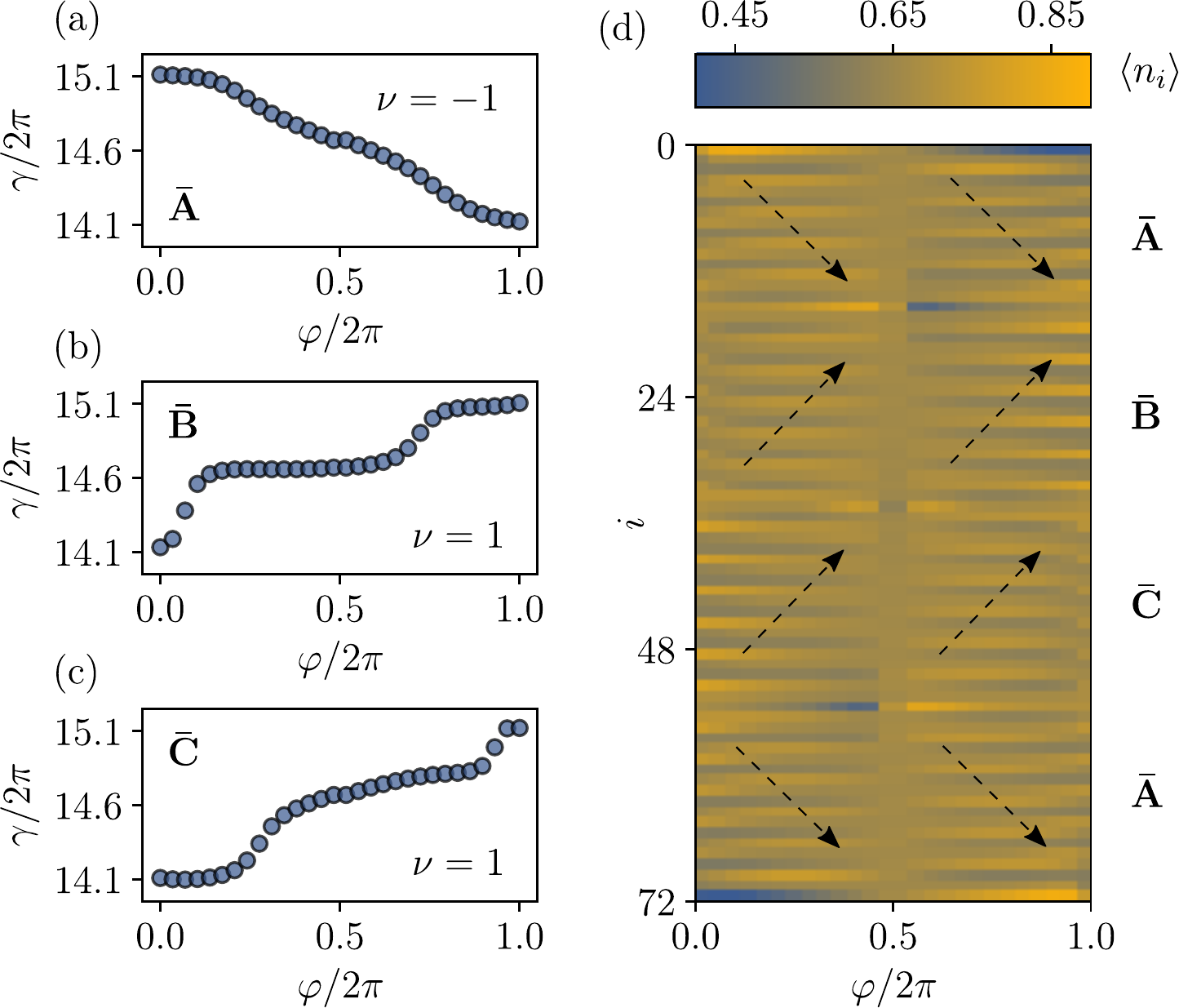}

\caption{\label{fig:chern_trimer} \textbf{Inter-soliton pumping with $\mathbb{Z}_4$ solitons: } {\bf (a)}, {\bf (b)} and {\bf (c)} show the local Berry phase $\gamma$ calculated at the middle of a finite chain with $L=42$ sites  as a function of the pumping parameter $\varphi$. They correspond to $\rho = 2/3$, but starting from the homogeneous configurations $\bar{\rm A}$, $\bar{\rm B}$ and $\bar{\rm C}$, respectively. Using eq.~\eqref{eq:chern_berry} we obtain Chern numbers equal to {\bf (a)} $\nu_{\bar{\rm A}}=-1$, {\bf (b)} $\nu_{\bar{\rm B}}=1$ and {\bf (c)} $\nu_{\bar{\rm C}}=1$. {\bf (d)} Real space bosonic occupation $\langle n_i \rangle$ through the pumping cycle of a system with three domain walls, starting from the $\bar{\rm A}\bar{\rm B}\bar{\rm C}\bar{\rm A}$ configuration. At the edges of the system, one particle is transported during the cycle, since the former act as boundaries between the region $\bar{\rm A}$ and the vacuum, with a the difference in Chern numbers of $-1$. Two of the defects separate regions where this difference is $\pm 2$, and the charge pumped through these defects is two, while it is zero across the trivial soliton (center). The direction of the transport (dashed arrows) depends on the sign of the Chern number, with $\Delta n = -\nu$.}
\end{figure}

Let us now move away from the single-particle scenario, and explore the pumping of bosons between the topological solitons in the strongly-correlated $\mathbb{Z}_2$BHM. We shall  use this pumping to discuss a generalised bound-defect correspondence that shines light on the topological origin of the bound fractional bosons. In this case, we implement the  adiabatic pumping by modifying the $\mathbb{Z}_2$BHM~\eqref{eq:z2_bhm}
 as follows
\begin{equation}
\label{eq:z2bhm}
\begin{aligned}
H =& -\sum_i \left[ b_i^\dagger(t+\alpha\sigma^z_{i,i+1})b_{i+1}^{\phantom{\dagger}}+\text{H.c.}\right] + \frac{U}{2}\sum_i n_i(n_i-1)\\
&+\sum_i\frac{\Delta_{\varphi,i}}{2}\sigma^z_{i,i+1}+\sum_i\beta_{\varphi,i}\sigma^x_{i,i+1},
\end{aligned}
\end{equation}
where $\Delta_{\varphi,i} = 2(-1)^i\delta\,\text{cos} (\varphi)$ and $\beta_{\varphi,i} = (-1)^i\delta\,\text{sin} (\varphi)$. By choosing $\delta \gg t$, we guarantee that the spins rotate periodically, effectively generating a superlattice modulation similar to the one we considered in the non-interacting Rice-Mele model (\ref{eq:H_dimer_pumping}). In practice, it is enough to fix $\delta = t$.

In this case, one must consider the many-body ground-state due to the presence of interactions. Therefore, the notion of the Chern number of the band must  be generalized to that of an  interacting system. The latter can be done by defining the Chern number in a 2D finite size system with the help of twisted boundary conditions~\cite{niu1985}. Alternatively, the Chern number can also be computed for a cylinder as the change of the Berry phase $\gamma(\varphi)$ during the pumping cycle~\cite{ent_spectrum_berry,ent_spectrum_berry_2}
\begin{equation}
\label{eq:chern_berry}
\nu = \frac{1}{2\pi}\int_0^{2\pi}\text{d}\varphi\,\partial_\varphi \gamma (\varphi).
\end{equation}
  We use the latter definition to infer the Chern number and characterize the topology of our pumping cycle.

We first consider the case of solitons around half filling. Figure~\ref{fig:chern_entanglement}{\bf (a)} shows the change in the many-body Berry phase during the pumping, which starts from the configuration B$=\bar{\rm A}$ at $\varphi=0$. We note that the many-body Chern number can be calculated from the finite changes of this quantity, which yield $\nu_{\bar{\rm A}}\approx-1$. The Chern number associated to the reversed pumping starting from the configuration A is $\nu_{{\rm A}}\approx1$, as shown in Fig.~\ref{fig:chern_entanglement}{\bf (b)}. These results allow us to draw an alternative picture of the topological origin of the fractionalized bound quasi-particles that appear in the $\mathbb{Z}_2$-BHM above/below  half filling. Similarly to the non-interacting case, the fractional bosons bound to the defects can be understood as remnants of the conducting states localized  at the 1D cylindrical interfaces that separate synthetic 2D regions with different Chern numbers. The difference between the Chern numbers of these regions predicts the number of bound modes, and its sign is related to the direction of the particle flow~\cite{defect_Hofstadter}. Figure~\ref{fig:chern_entanglement}{\bf (c)} shows how the real-space bosonic occupation evolves for a finite chain with two domain walls.  This figure clearly supports the above prediction: as a consequence of the different topological invariants, quantised charge is transported between the two existing solitons. The latter separate regions with different Chern numbers in the extended 2D system, and the direction of the transport is different for each one.

As stated above, this pumping offers an alternative take on the topological origin of the fractionalized bosons. However, for the half-filled case, it is not essential as the topological solitons always separate regions with a different Berry phase, and one can develop a topological characterization based solely on the equilibrium properties of the1D model.
Let us now move the case of two-third filling, where solitons interpolate between regions with the same Berry phase, and pumping becomes essential to unravel the topological aspects of the fractionalized bosons.  We first compute the Chern numbers associated to the three degenerate configurations ${\rm \bar{A}}$, ${\rm \bar{B}}$ and ${\rm \bar{C}}$. Figures~\ref{fig:chern_trimer} {\bf (a)}, {\bf (b)} and {\bf (c)} show the Berry phases for the three configurations ${\rm \bar{A}}$, ${\rm \bar{B}}$ and ${\rm \bar{C}}$, the change of which as a function of the pumping phase leads to  Chern numbers $\nu_{\rm \bar{A}}=-1$, $\nu_{\rm \bar{B}}=1$ and $\nu_{\rm \bar{C}}=1$ respectively. Two configurations have therefore the same Chern number, which means that one of the solitons  does not host a topologically-protected conducting state in the extended 2D system. As a consequence, as can be observed in Fig.~\ref{fig:pumping_dimer}{\bf (c)} for the  equivalent situation with free fermions,  the  in-gap state of the corresponding defect does not connect the valence and conduction  bands, and can be thus removed without closing the gap.

In the full many-body  $\mathbb{Z}_2$BHM~\eqref{eq:z2bhm}, the aforementioned absence of spectral flow means  that the ${\rm \bar{B}}{\rm \bar{C}}$ soliton will not accumulate an integer pumped charge during the adiabatic cycle. In Figure~\ref{fig:chern_trimer}{\bf (d)}, we represent the time evolution of the bosonic density through the cycle for a system with three solitons. It can be observed how the bosonic density is pumped along distinct directions in the the three different regions, and that no particle number is being pumped in one of the defects, which is consistent with the above argument. In the contrary, for the other  ${\rm \bar{A}}{\rm \bar{B}}$  and  ${\rm \bar{C}}{\rm \bar{A}}$ solitons, the extended interfaces have a different Chern number, and the quantization of the adiabatic pumping can be thus used for the topological characterization of the SPTd. Let us highlight once more, that the Berry phase of all these composite defects are all the same $\gamma_{\rm \bar{A}}=\gamma_{\rm \bar{B}}=\gamma_{\rm \bar{C}}=\pi$. Therefore, although the fractionalized bound states do not arise in the interface of two topologically distinct regions in the equilibrium situation, we see that the extended regions via the pumping do indeed interpolate between regions with a different Chern number, guaranteeing thus the topological robustness of the pumping between these SPT-d.

 As a summary, the pumping argument has allowed us to uncover the topological origin of the localized bosonic states associated to the $\mathbb{Z}_4$ solitons. As opposed to the case of $\mathbb{Z}_2$ solitons, in the later this origin could not be inferred from the topological properties of the different SSB sectors in 1D, as they all belong to the same topological phase. We note that, in these simulations,  we have pinned down the defects since everything is computed in the ground state of the system, which coincides with the time-dependent calculation for an adiabatic evolution. In a cold-atom experiment, however, the defects are expected to move during the cycle. The results for the transported charge, however, should hold also in this situation.

\section{\bf Conclusions and Outlook}
\label{sec:conclusions}

In this work, we have explored the existence of symmetry-protected topological defects in intertwined topological phases, where both spontaneous symmetry breaking and topological symmetry protection cooperate, giving rise to exotic states of matter. We have analyzed how the topological solitons, which interpolate between different symmetry-broken sectors, can host fractionalized matter quasi-particles, and how the back action of the matter sector on the topological solitons is crucial to encounter this phenomenon directly in the groundstate. We have presented a thorough analysis of the $\mathbb{Z}_2$ Bose-Hubbard model, where such interesting effects appear via a bosonic version of the Peielrs' mechanism. This leads to different types of $\mathbb{Z}_n$ solitons, showing  distinct  fractionalized bosonic densities bound to the solitons, and organized according to a rich variety of  layouts: from pinned configurations with few topological defects, to solitonic lattices with corresponding crystalline densities of  fractionalized bosons. 

The remarkable progress in cold-atom quantum simulators, together with the recent interest in bosonic lattice models with $\mathbb{Z}_2$-dressed tunnelling, points to a very promising avenue for the realization of the  $\mathbb{Z}_2$ Bose-Hubbard model. As conveyed in this work, this experiment would allow for the first direct observation of dynamic topological solitons with bound fractionalized quasi-particles, overcoming past difficulties in conjugate-polymer science, and giving novel insights in genuinely quantum-mechanical topological defects. 

In addition to the aforementioned  static phenomena, we have also explored the quantization of particle transport in adiabatic pumping protocols. As a consequence of the topological solitons and bound quasi-particles, we have observed that an integer number of bosons can be pumped between the topological solitons, and that this phenomenon can be used to derive a generalised bulk-defect correspondence. Given the variety of intertwined topological phases in the   $\mathbb{Z}_2$ Bose-Hubbard model, it turns out that one can find fractionalized quasi-particles bound to solitons that separate regions with the same topological Berry phase. It is only through the adiabatic inter-soliton pumping of bosons, and the connection to extended 2D system with the adiabatic parameter playing the role of an extra synthetic dimension, that the topological origin of the bound quasi-particles can be neatly understood. We have shown that, despite appearing at the interface of SSB sectors with the same topological Berry phase,  these  quasi-particles bound to the solitons can be understood, via the pumping scheme,  as remnants of protected edge states in 2D that are bound to the boundaries separating synthetic regions with different Chern numbers. This generalised bulk-defect correspondence thus clarifies the topological origin of these  quasi-particles via a Kaluza-Klein dimension reduction.

So far, our discussion has focused on the topological features of the bosonic sector at zero temperature, which are allowed by their coupling to the  solitons in the spin sector. As an outlook, we would like to note that, at finite temperatures, some additional fractionalization can also take place in the spin solitons, which reminds of the exotic properties of the spinon branches in antiferromagnetic Heisenberg chains. In the regime of large Peierls-Nabarro barriers, the existence of two soliton branches dicussed in Sec.~\ref{subsec:z2} has a direct consequence: the domain walls develop a fractional value of the statistical-interaction parameter $g=1/2$, which leads to fractional exclusion statistics  with characteristic semionic thermodynamics~\cite{FES_Haldane}. At other fractional fillings, such as $\rho=2/3$ of Sec.~\ref{subsec:z4}, we expect that the statistical interaction will be 
$g=1/3$ in the regime of large Peierls-Nabarro barriers. This suggests an interesting connection to the groundstate degeneracy. In the future, we will explore these questions in detail, and study the quantitative thermodynamical consequences that they may give rise to. 

\acknowledgements

This project has received funding from the European Union’s Horizon 2020 research and innovation programme under the Marie Sk\l{}odowska-Curie grant agreement No 665884, the Spanish Ministry MINECO (National Plan 15 Grant: FISICATEAMO No. FIS2016-79508-P, SEVERO OCHOA No. SEV-2015-0522, FPI), European Social Fund, Fundaci\'{o} Cellex, Generalitat de Catalunya (AGAUR Grant No. 2017 SGR 1341, CERCA/Program), ERC AdG NOQIA, EU FEDER, MINECO-EU QUANTERA MAQS, and the National Science Centre, Poland-Symfonia Grant No. 2016/20/W/ST4/00314. A. D. is financed by a Juan de la Cierva fellowship (IJCI-2017-33180). A.B. acknowledges support from the Ram\'on y Cajal program RYC-2016-20066, and CAM/FEDER Project S2018/TCS- 4342 (QUITEMAD-CM).

\bibliographystyle{apsrev4-1}
\bibliography{bibliography}

\end{document}